\documentclass[aps, prd, twocolumn, floatfix, superscriptaddress, nofootinbib]{revtex4-1}

\usepackage{amsmath, amssymb, physics}
\usepackage{microtype}
\usepackage[T1]{fontenc}
\usepackage[utf8]{inputenc}

\usepackage{graphicx}
\usepackage[caption=false]{subfig}

\usepackage{multirow}

\usepackage{xcolor}
\definecolor{linkcolor}{rgb}{0.0, 0.3, 0.5}
\usepackage[colorlinks, linkcolor=linkcolor, citecolor=linkcolor, urlcolor=linkcolor]{hyperref}

\newcommand{\Cornell}{Center for Astrophysics and Planetary Science,
    Cornell University, Ithaca, New York 14853, USA}
\newcommand{\Caltech}{Theoretical Astrophysics 350-17,
    California Institute of Technology, Pasadena, CA 91125, USA}

\newcommand{\dii}{{\tt I1}}
\newcommand{\diii}{{\tt I2}}
\newcommand{\diir}{{\tt I1R}}

\newcommand{\dbi}{{\tt B1}}
\newcommand{\dbii}{{\tt B2}}
\newcommand{\dbir}{{\tt B1R}}

\newcommand{\dfv}{{\tt FV}}
\newcommand{\dfvr}{{\tt FVR}}

\begin{document}

\title{
General-relativistic neutron star evolutions with the discontinuous Galerkin method
}

\author{Fran\c{c}ois H\'{e}bert}
\email{fhebert@caltech.edu}
\affiliation{\Cornell}
\affiliation{\Caltech}

\author{Lawrence E.\ Kidder}
\affiliation{\Cornell}

\author{Saul A.\ Teukolsky}
\affiliation{\Cornell}
\affiliation{\Caltech}

\date{\today}

\begin{abstract}
Simulations of relativistic hydrodynamics often need both high accuracy and
robust shock-handling properties. The discontinuous Galerkin method combines
these features --- a high order of convergence in regions where the solution is
smooth and shock-capturing properties for regions where it is not --- with
geometric flexibility and is therefore well suited to solve the partial
differential equations describing astrophysical scenarios. We present here
evolutions of a general-relativistic neutron star with the discontinuous
Galerkin method. In these simulations, we simultaneously evolve the spacetime
geometry and the matter on the same computational grid, which we conform to the
spherical geometry of the problem. To verify the correctness of our
implementation, we perform standard convergence and shock tests. We then show
results for evolving, in three dimensions, a Kerr black hole; a neutron star in
the Cowling approximation (holding the spacetime metric fixed); and, finally, a
neutron star where the spacetime and matter are both dynamical. The evolutions
show long-term stability, good accuracy, and an improved rate of convergence
versus a comparable-resolution finite-volume method.
\end{abstract}

\maketitle

\section{Introduction}
\label{sec:introduction}

Numerical simulations are a crucial tool in the study of core-collapse
supernovae, compact binary mergers, accretion disks with relativistic jets, and
other energetic astrophysical sources. In these events, the dynamics are
governed by the high-density matter and its coupling to the strong
gravitational field. Nuclear reactions, neutrino physics, and magnetic fields
can also play significant roles. Because of the highly nonlinear nature of the
underlying general-relativistic hydrodynamics (GR-hydro), simulations are
necessary to obtain observable predictions from physics models. Achieving
sufficient accuracy in the simulation outputs (e.g., gravitational waveforms,
ejected masses, and nucleosynthesis products) remains a challenge, however.
High resolution is needed to resolve multiscale fluid flows, and the presence
of shocks in the matter reduces the accuracy of the numerical schemes.

The standard approach taken in present-day GR-hydro codes is to cast the
partial differential equations into conservative form and discretize them using
a finite-volume (FV) method (see reviews \cite{lrr-2008-7, Marti2015,
Balsara2017:FVMs} for an overview and history). FV methods are favored for
their robustness and the various ``shock-capturing'' schemes that enable them
to handle fluid shocks and stellar surfaces. The Einstein equations for the
spacetime geometry are typically solved with a finite-difference method on the
same grid but can instead be solved with a spectral method on a different
computational grid \cite{Duez:2008rb}. Over the past decade, the application of
improved high-resolution shock-capturing schemes (e.g., the piecewise parabolic
method (PPM) \cite{PPM, Mignone2005} and the weighted essentially
non-oscillatory (WENO) scheme \cite{Jiang1996202}) and higher-order difference
schemes has led to significant advances in the accuracy and stability of the
numerical results (e.g., for core-collapse supernovae \cite{bruenn:13,
moesta:15}, binary mergers \cite{Kiuchi2014, Bernuzzi2016, FoucartM1:2016}, and
accretion flows \cite{2014MNRAS.441.3177M, Porth2017:BHAC, Kelly2017}). In
spite of these successes, FV methods have inherent limitations when used as
high-order methods: the large stencils required for the differencing and
shock-capturing schemes make it difficult to adapt the grid to the problem
geometry and can also lead to challenges in efficiently parallelizing the
algorithm.

In the pursuit of improved accuracy and efficiency, discontinuous Galerkin (DG)
methods have recently emerged as a promising contender for astrophysical
problems. DG methods share properties with both spectral methods and FV methods
--- they inherit the high-order accuracy of the former for smooth solutions
while maintaining the robust shock-handling properties of the latter. They are
geometrically flexible, enabling the use of grids adapted to the problem
geometry. They are well suited to $hp$-adaptivity, where the grid resolution
can be set either by adjusting the order of the polynomial approximation within
an element ($p$-refinement) or by adjusting the size of the element
($h$-refinement). Finally, DG methods are locally formulated, enabling
efficient parallelization and good scaling.

The application of DG methods to problems in relativistic astrophysics is
recent and remains exploratory in nature. With several of these explorations
focusing on the evolution of the spacetime geometry, different formulations of
Einstein's equations have been investigated. In an early study, Zumbush
\cite{zumbusch2009} obtained a space-time DG scheme for the linearized vacuum
Einstein equations in harmonic gauge. For the commonly used
Baumgarte-Shapiro-Shibata-Nakamura (BSSN) formulation of the Einstein
equations, Field \emph{et al.}\ \cite{field:10} and Brown \emph{et al.}\
\cite{brown:12dg} developed DG schemes in spherical symmetry. More recently,
Miller and Schnetter \cite{Miller:2016vik} developed an operator-based (vs the
typical differential equation-based) DG discretization of the BSSN equations
and showed success in evolving three-dimensional (3D) test problems. Using a
new first-order form of the constraint-damping Z4 formulation (FO-CCZ4),
Dumbser \emph{et al.}\ \cite{Dumbser2017:FOCCZ4} evolved a single black hole
(BH) spacetime using a puncture and showed a short-timescale ``proof of
concept'' evolution of a binary BH system.

Efforts on the hydrodynamics side began with Radice and Rezzolla
\cite{Radice:2011qr}, who presented a formulation of DG for the evolution of
fluids in curved spacetimes and evolved a neutron star (NS) in spherical
symmetry. In their work, the spacetime is treated self-consistently by
satisfying a radial constraint equation. In Ref.\ \cite{Zhao2013}, Zhao and
Tang implemented DG with a WENO shock-capturing scheme for special-relativistic
hydrodynamics in one and two dimensions. Bugner \emph{et al.}\
\cite{Bugner:2015gqa} were the first to apply DG to a 3D astrophysical fluid
problem, evolving a NS in the Cowling approximation (i.e., fixed background
metric). In a DG code using a task-based parallelism paradigm
(\textsc{SpECTRE}), Kidder \emph{et al}.\ \cite{kidder:16} showed
special-relativistic magnetohydrodynamics tests in two and three dimensions.
Anninos \emph{et al}.\ \cite{Anninos2017:CosmosDG} and Fambri \emph{et al}.\
\cite{Fambri2018} (see also Ref.\ \cite{Koppel2017}) implemented DG schemes
with adaptive mesh refinement (AMR) for applications to special- and
(fixed-background) general-relativistic magnetohydrodynamics, and showed
results in two and three dimensions.

In this paper, we use a DG method to evolve a NS in coupled GR-hydro in three
dimensions (prior efforts in this direction are the subject of theses by
H{\'e}bert \cite{HebertThesis} and Bugner \cite{BugnerThesis}). As tests of
our implementation, we also evolve a NS in the Cowling approximation and a Kerr
BH. In these simulations, we investigate the use of cubed-sphere grids
conforming to the spherical geometry of the BH and NS problems. We adopt the DG
formulation described by Teukolsky \cite{teukolsky2015}, using the generalized
harmonic formulation of Einstein's equations
\cite{Friedrich1985,Pretorius2005c,Gundlach2005} and the Val\`{e}ncia
formulation \cite{lrr-2008-7} of the general-relativistic hydrodynamics.

We implement our DG code in the framework of the Spectral Einstein Code
\cite{SpECwebsite} (\textsc{SpEC}). \textsc{SpEC} combines a multidomain
penalty spectral method to evolve binary BH spacetimes
\cite{Scheel2006,Szilagyi:2009qz, Hemberger:2012jz} with a FV method to evolve
the matter in BH-NS \cite{Duez:2008rb} and NS-NS
\cite{FoucartM1:2016,Haas:2016} systems. Our DG GR-hydro code is independent
from \textsc{SpEC}'s FV component and is instead built on the algorithms from
\textsc{SpEC}'s vacuum spectral code: spectral bases and differentiations,
domain mappings, communication, etc.

There are two main goals of this work:
\begin{enumerate}
\item Explore the DG method as a means of solving the GR and hydrodynamics
  equations simultaneously. As we will see below, the equations of the two
  theories take fundamentally different forms (conservative vs
  nonconservative), so it is not \emph{a priori} obvious that solving them on
  the same grid with the same technique will work.
\item Explore the use of conforming grids for BH and NS applications. In these
  grids, cubical elements are mapped to match the spherical geometry of an
  excision boundary inside the BH or the spherical boundary at large distances
  from the BH or NS.
\end{enumerate}

This paper is organized as follows. We first summarize the formulation of our
DG method in Sec.\ \ref{sec:dg}. We discuss our use of geometrically adapted
grids, ``manual'' mesh refinement, and limiters in Sec.\ \ref{sec:philosophy}.
We detail the GR-hydro equations and associated algorithms of our numerical
implementation in Sec.\ \ref{sec:grhydro}. To validate our code, we perform
standard test cases; we show these in Sec.\ \ref{sec:tests}. We present our
results --- NS evolutions using the DG method --- in Sec.\ \ref{sec:results},
before concluding in Sec.\ \ref{sec:conclusion}.

\section{Discontinuous Galerkin formulation}
\label{sec:dg}

Our code uses a DG method to solve conservation laws in curved spacetimes and
also to evolve the spacetime itself. We express the spacetime metric
$g_{\mu\nu}$ using the standard 3+1 form
\begin{align}
ds^2 &= g_{\mu\nu}dx^{\mu}dx^{\nu}
\nonumber
\\
&= - \alpha^2 dt^2 + \gamma_{ab}(dx^a + \beta^a dt)(dx^b + \beta^b dt),
\end{align}
where $\alpha$ is the lapse function, $\beta^a$ is the shift vector, and
$\gamma_{ab}$ is the spatial metric (with determinant $\gamma$) on
hypersurfaces of constant time $t$. Our index convention is as follows. Greek
indices ($\mu$,$\nu$,...) refer to spacetime components and range from 0 to $d$
in $d$ spatial dimensions. Latin indices ($a$,$b$,...) refer to spatial
components and range from 1 to $d$. Repeated indices are summed over. We denote
by $\textbf{x}$ the spatial point with coordinates $x^a$. We use units where
$G,c=1$. We additionally set $M_{\odot}=1$ for the NS simulations in Sec.
\ref{sec:results}.

A conservation law in this curved spacetime can be written as a 4-divergence
$\nabla_{\mu} F^{\mu} = s$, where $\nabla_{\mu}$ is the covariant derivative,
$F^{\mu}$ encodes the conserved quantity $u = F^0$ and its corresponding
spatial flux vector $F^a(u)$, and $s$ is the source term for $u$. Separating
the time and spatial components gives the more common form
\begin{equation}
\label{eq:conslaw}
\frac{1}{\sqrt{\gamma}}\partial_t(\sqrt{\gamma} u)
+ \frac{1}{\sqrt{\gamma}}\partial_a(\sqrt{\gamma} F^a) = s,
\end{equation}
which we aim to solve for $\sqrt{\gamma} u(\textbf{x},t)$ given initial
conditions $\sqrt{\gamma} u(\textbf{x},0)$ and suitable boundary conditions.
When solving a system of conservation laws (e.g., for mass, energy, and
momentum in hydrodynamics), $u$ is a vector of several conserved quantities,
and $F^a$ is a vector of flux vectors.

We numerically solve the conservation law%
\footnote{The conservation law is discretized (see Sec. \ref{sec:dgcons}) and
solved for a numerical approximation to the true solution $u$. We do not
make the distinction between the approximate and true solutions.}
using a strong-form, nodal DG method on square/cube elements. In this section,
we summarize the method and give the specifics of our implementation. We follow
the formulation given by Teukolsky in Ref.\ \cite{teukolsky2015}, in which
greater detail may be found.

\subsection{Representing the solution}

We divide the spatial domain into $K$ elements. On each element, we expand the
quantities $u$, $F^a$, $s$, etc., over a set of polynomial basis functions
$\phi_i$, e.g.,
\begin{equation}
\label{eq:nodalexp}
u(\textbf{x},t) = \sum_i u_i(t) \phi_i(\textbf{x}).
\end{equation}
We adopt a nodal representation: we evolve the values
$u_i(t) = u(\textbf{x}_i,t)$ at the nodes $\textbf{x}_i$ of the computational
grid, and the $\phi_i$ interpolate between these grid nodes. Below, we define
these quantities; more detailed discussion can be found in textbooks
\cite{Hesthaven2008, kopriva2009implementing}.

The partition into elements is chosen so that each element is a mapping of a
topologically simple reference element: a cube (in three dimensions), square
(in two dimensions), or interval (in one dimension). The mapping from the
reference element coordinates $\bar{\textbf{x}}$ to the computational
coordinates $\textbf{x} = \textbf{x}(\bar{\textbf{x}})$ of each element has a
Jacobian matrix
\begin{equation}
\textbf{J} = \frac{\partial x^a}{\partial x^{\bar{a}}}
\end{equation}
and Jacobian $J = \det \textbf{J}$.

In each direction, the $x^{\bar{a}}$ coordinate spans the interval $[-1,1]$,
and on this interval, we place the nodes $x^{\bar{a}}_i$ of a
Gauss-Legendre-Lobatto quadrature. The one-dimensional (1D) Lagrange
interpolation polynomials $\ell_j(x^{\bar{a}})$ are defined on these nodes and
satisfy $\ell_j(x^{\bar{a}}_i) = \delta_{ij}$. In the full $d$ dimensions, we
construct a tensor-product grid --- we obtain the grid nodes
$\bar{\textbf{x}}_i$ from the direct product of the $x^{\bar{a}}_i$ and the
basis functions $\phi_i$ from the product of the $\ell_i(x^{\bar{a}})$, e.g.,
(with some abuse of indices to indicate the tensor product)
\begin{equation}
\phi_i(\bar{\textbf{x}}) \to \phi_{ijk}(\bar{\textbf{x}})
= \ell_i(x^{\bar{1}}) \ell_j(x^{\bar{2}}) \ell_k(x^{\bar{3}}).
\end{equation}
With $N_p$ nodes in the $x^{\bar{a}}$ coordinate, $\ell_i(x^{\bar{a}})$ is a
polynomial of degree $N = N_p - 1$. When $N$ is the same in all directions, we
say we have an $N^{\text{th}}$-order DG element.

We will occasionally use a modal representation in which the solution is
expanded over a basis of orthonormal polynomials, e.g.,
\begin{equation}
u(\bar{\textbf{x}},t) = \sum_i \hat{u}_i(t) \psi_i(\bar{\textbf{x}}).
\end{equation}
The $\hat{u}_i$ are the expansion weights, and the $\psi_i$ are obtained from
the tensor product of orthonormal 1D basis functions, the Legendre polynomials
$P_l$. The Vandermonde matrix $\mathcal{V}_{ij} = P_j(x_i)$ gives the
transformation between the nodal and modal representations,
\begin{equation}
u_i = \sum_j \mathcal{V}_{ij} \hat{u}_j.
\end{equation}

\subsection{DG for conservation laws}
\label{sec:dgcons}

We impose the conservation law \eqref{eq:conslaw} in a Galerkin sense, by
integrating the equation against each basis function $\phi_i$ on each element.
We integrate over proper volume $\sqrt{\gamma} d^3x$, giving
\begin{equation}
\int \left[ \partial_t(\sqrt{\gamma} u) + \partial_a(\sqrt{\gamma} F^a) -
  \sqrt{\gamma} s \right] \phi_i(\textbf{x}) d^3 x = 0.
\end{equation}
To establish the flow of information between neighboring elements, we integrate
the flux divergence term by parts, and apply Gauss's law to the resulting
boundary term (see Ref.\ \cite{teukolsky2015}),
\begin{equation}
\label{eq:dgstep1}
\begin{split}
\int \partial_a(\sqrt{\gamma} F^a) \phi_i(\textbf{x}) d^3 x
=
- \int \sqrt{\gamma} F^a \partial_a \phi_i(\textbf{x}) d^3 x
\\
+ \oint F^a n_a \phi_i(\textbf{x}) d^2 \Sigma.
\end{split}
\end{equation}
Here, $d^2\Sigma$ is the proper surface element on the element's boundary, and
$n_a$ is the outward-directed unit normal.

The flux vector $F^a$ is double valued on the boundary because of the local
(i.e., discontinuous) nature of the solution. However, for the scheme to be
conservative, a unique flux must cross the boundary between two adjacent
elements --- this is the so-called \textit{numerical} flux $F^{a*}$. The
numerical flux is computed from the data on both sides of the boundary and so
requires the communication of boundary data between nearest-neighbor elements.
We substitute $F^a \to F^{a*}$ in the last term of \eqref{eq:dgstep1}.

We now undo the integration by parts, using \eqref{eq:dgstep1} to eliminate the
second (i.e., $\partial_a \phi_i$) term (this time, however, we do not
substitute in the numerical flux) and obtain
\begin{equation}
\begin{split}
\int \partial_a(\sqrt{\gamma} F^a) \phi_i(\textbf{x}) d^3 x
\to
\int \partial_a(\sqrt{\gamma} F^a) \phi_i(\textbf{x}) d^3 x
\\
+ \oint (F^{a*} - F^a) n_a \phi_i(\textbf{x}) d^2 \Sigma.
\end{split}
\end{equation}
The surface integral term provides a boundary condition on the element and
serves to connect the solution between neighboring elements of the domain.
Defining $F = (F^{a*} - F^a) n_a$ and putting the terms back together, we get
the DG equation in integral form,
\begin{equation}
\label{eq:dgint}
\begin{split}
\int \left[
  \partial_t(\sqrt{\gamma} u) + \partial_a(\sqrt{\gamma} F^a) - \sqrt{\gamma} s
 \right] \phi_i(\textbf{x}) d^3 x = \\
 - \oint F \phi_i(\textbf{x}) d^2 \Sigma.
\end{split}
\end{equation}

To obtain a form more suitable for computation, we first expand each term of
\eqref{eq:dgint} using the nodal expansion \eqref{eq:nodalexp}. We rewrite the
integrals in the reference coordinates $\bar{\textbf{x}}$, where
$d^3x \to J d^3\bar{x}$ and $d^2\Sigma \to \sqrt{^{(2)}\gamma} d^2\bar{x}$,
with $^{(2)}\gamma$ the determinant of the two-dimensional (2D) metric induced
by $\gamma_{ab}$ on the surface. Finally, we evaluate the integrals with a
Gauss-Lobatto quadrature rule. By using the grid nodes $\bar{\textbf{x}}_i$ as
the quadrature nodes, we can use the identity
$\ell_i(x^{\bar{1}}_j)=\delta_{ij}$ to greatly simplify the scheme. The
tradeoff is that the quadrature rule will not be exact --- especially when a
nontrivial Jacobian $J$ multiplies the integrand --- and this can lead to
aliasing and introduce numerical instabilities that require filtering.

Finally, after simplifying the geometric factors on the boundary terms (see
Ref.\ \cite{teukolsky2015}, Appendix A) and dividing through by common factors,
we arrive at the evolution equation,
\begin{widetext}
\begin{align}
\label{eq:dgfinal}
\frac{d(\sqrt{\gamma}u)_{ijk}}{dt}
&+
\Big[
\frac{\partial x^{\bar 1}}{\partial x^a}\Big|_{ijk}\sum_l D_{il}^{\bar 1}
 \left(\sqrt{\gamma}F^a\right)_{ljk}
+
\frac{\partial x^{\bar 2}}{\partial x^a}\Big|_{ijk}\sum_m D_{jm}^{\bar 2}
 \left(\sqrt{\gamma}F^a\right)_{imk}
+
\frac{\partial x^{\bar 3}}{\partial x^a}\Big|_{ijk}\sum_n D_{kn}^{\bar 3}
 \left(\sqrt{\gamma}F^a\right)_{ijn}
\Big]
-
(\sqrt{\gamma}s)_{ijk}
\nonumber
\\ &=
-
\frac{\sqrt{\gamma^{\bar{1}\bar{1}}_{Njk}}}{w_N}(\sqrt{\gamma}F)_{Njk}
  \delta_{iN}
-
\frac{\sqrt{\gamma^{\bar{2}\bar{2}}_{iNk}}}{w_N}(\sqrt{\gamma}F)_{iNk}
  \delta_{jN}
-
\frac{\sqrt{\gamma^{\bar{3}\bar{3}}_{ijN}}}{w_N}(\sqrt{\gamma}F)_{ijN}
  \delta_{kN}
\nonumber
\\& \phantom{=}
+
\frac{\sqrt{\gamma^{\bar{1}\bar{1}}_{0jk}}}{w_0}(\sqrt{\gamma}F)_{0jk}
  \delta_{i0}
+
\frac{\sqrt{\gamma^{\bar{2}\bar{2}}_{i0k}}}{w_0}(\sqrt{\gamma}F)_{i0k}
  \delta_{j0}
+
\frac{\sqrt{\gamma^{\bar{3}\bar{3}}_{ij0}}}{w_0}(\sqrt{\gamma}F)_{ij0}
  \delta_{k0}.
\end{align}
\end{widetext}
Here, $D^{\bar 1}_{il}$ is the differentiation matrix along the $x^{\bar{1}}$
direction, given by
\begin{equation}
D_{il}^{\bar 1}=\partial_{\bar 1} \ell_l\big(x^{\bar 1}\big)\big|_i.
\end{equation}

Although our derivation and resulting evolution equation \eqref{eq:dgfinal} are
given for the 3D case, restricting to a lower-dimensional problem is
straightforward. For instance, in a 2D problem, the third tensor-product index
on each term is dropped (e.g., $u_{ijk} \to u_{ij}$), as are the $\bar{3}$
terms of the flux derivative and flux boundary terms.

\subsection{DG for the Einstein equations}
\label{sec:dghyp}

We use a formulation of the Einstein equations, detailed in the next section,
that cannot be written in conservative form. These equations are instead in
hyperbolic form,
\begin{equation}
\partial_t u + A^a \partial_a u = s,
\end{equation}
where the matrices $A^a$ and the vector $s$ may be functions of $u$, but not of
derivatives of $u$. To obtain the corresponding DG algorithm, we again multiply
by a basis function $\phi_i$ and integrate over the proper volume element. We
integrate by parts twice, substituting the numerical flux after the first
integration, to obtain the integral form akin to
\eqref{eq:dgint},
\begin{equation}
\begin{split}
\int \left[ \partial_t u + A^a \partial_a u - s \right] \phi_i(\textbf{x})
\sqrt{\gamma} d^3 x = \\
 - \oint \left[ (A^a u)^* - (A^a u) \right] n_a \phi_i(\textbf{x}) d^2 \Sigma.
\end{split}
\end{equation}
Evaluating the integrals as before, we find
\begin{align}
\label{eq:dgfosh}
\frac{du_{ijk}}{dt}
&+
A^a_{ijk} \Big[
\frac{\partial x^{\bar 1}}{\partial x^a}\Big|_{ijk}\sum_l D_{il}^{\bar 1}
  u_{ljk}
+ ...
\Big]
- s_{ijk}\nonumber\\
&=
-
\frac{\sqrt{\gamma^{\bar{1}\bar{1}}_{Njk}}}{w_N} \big([(A^a u)^*
  - (A^a u)]n_a\big)_{Njk} \delta_{iN}
+ ...\ .
\end{align}
This result is analogous to \eqref{eq:dgfinal}, so we have reproduced here only
one term of each type.

\subsubsection{Comparison with \textsc{SpEC}'s penalty spectral algorithm}

\textsc{SpEC} solves the Einstein equations using a multidomain penalty
pseudospectral method (see, e.g., Ref.\ \cite{Hesthaven2007}). This method is
closely related to our nodal DG method: the DG boundary term represents a
particular type of penalty term, one chosen to enforce conservation via the
numerical flux. Indeed, the spectral method in \textsc{SpEC} takes the form of
\eqref{eq:dgfosh} with an upwind flux, differing only in the numerical
prefactor multiplying the boundary flux term. Where our DG method has a
prefactor of $1/w_N$ arising from the Legendre Gauss-Lobatto quadrature rule,
the \textsc{SpEC} penalty method instead uses the prefactor derived for
stability of a Chebyshev penalty method \cite{Gottlieb2001}. In numerical
experiments (not reported in this paper), we observe a higher order of
convergence under $h$-refinement from the DG method (order $N+1$) than from
\textsc{SpEC}'s spectral method (order $N$).

\section{Approach to grid structure, mesh refinement, and limiting}
\label{sec:philosophy}

Early applications of the DG method to problems in astrophysics have used
uniform grids. We adopt a different philosophy and take advantage of the DG
method's geometric flexibility to tailor our grid to the problem being solved.
This approach was also taken by Refs.\ \cite{Koppel2017, Anninos2017:CosmosDG},
which use a 2D wedge-shaped domain when evolving gas flows in a BH spacetime.
We discuss here our choice of grid structures, mesh refinement, and limiting.

\subsection{Grid structure and mesh refinement}
\label{sec:philosophygrid}

It is well known that constructing the computational grid to mirror the
underlying symmetries of the problem can greatly increase the accuracy of a
numerical method. In astrophysical problems, the symmetry is often spherical,
reflecting the gravitational potential of a star or BH. The use of a conforming
spherical grid comes with a loss of generality: the grid must remain centered
on the astrophysical body. This is especially important when taking advantage
of the spherical grid to excise the singularity inside a BH. With the use of
moving grids \cite{Scheel2006} and control systems \cite{Hemberger:2012jz},
however, conforming grids can be successfully used in simulations of binary
mergers.

The evolutions shown in this paper make use of two basic types of grid
structures:
\begin{enumerate}
\item Cartesian grids, obtained by a straightforward affine mapping (a
  translation and a scaling) of the reference element. These grids are used in
  several standard test problems.
\item Cubed-sphere grids, obtained by conforming several cube-like elements to
  the surface of a sphere, using mappings detailed in Appendix
  \ref{app:cubedsphere} and illustrated in, e.g., Fig.\ \ref{fig:bhdomains}.
  These grids are used for problems with spherical geometry such as single BH
  or NS evolutions. The cubed-sphere grid may cover a hollow spherical shell,
  allowing for excision of the spacetime region inside the BH's event horizon,
  or a filled ball, for evolution of the full NS. As we consider isolated
  systems at rest, moving grids are not needed.
\end{enumerate}

To further take advantage of the geometric flexibility of the DG method, we use
$hp$-adaptivity to vary the spatial resolution across the simulation domain.
The AMR infrastructure of \textsc{SpEC} is designed to operate under a
restricted set of conditions and is not general enough to handle the shocks and
surfaces encountered in the hydrodynamics evolutions. We instead manually set
up fixed mesh refinement, where we initially assign the size and order of the
DG elements based on \emph{a priori} knowledge of the solution. When
constructing the grid for a NS evolution, for instance, we use larger,
higher-order elements inside the star and smaller, lower-order elements at the
surface. We use ``higher-order'' (``lower-order'') as a qualitative description
of a DG element, typically referring to elements with $N \gtrsim 3$
($N \lesssim 2$).

The \textsc{SpEC} framework, designed and optimized for evolutions on
$\mathcal{O}(10\text{--}100)$ spectral elements, scales poorly to the large
number of elements often used in DG simulations. In spite of several
improvements to the data structures, we find that the code's memory usage and
parallelism become inefficient when the domain approaches $\mathcal{O}(10^4)$
elements. We therefore stay below this threshold in most of the tests
presented. This restriction on the maximum number of elements would be
problematic for a typical DG implementation, in which the domain is split into
a regular grid of many small cubical elements. As we instead conform our grids
to the problem geometry, we obtain satisfactory accuracy using many fewer
elements.

\subsection{Limiting}
\label{sec:philosophylimiter}

In DG elements containing a shock or surface in the fluid, the solution is
susceptible to spurious oscillations (Gibbs phenomenon) and overshoots. If
unaddressed, these overshoots can lead to unphysical fluid states (e.g.,
negative densities) in which the fluid equations are no longer solvable. A
limiter controls these oscillations and overshoots by modifying the solution in
a way that is conservative and --- ideally --- does not overly degrade the
accuracy of the method.

Typical DG implementations apply the limiter agnostically across the uniform
grid. A ``troubled-cell'' detector identifies cells containing spurious
oscillations and applies the limiter to those cells. While this is the most
general way to set up the problem, finding a general troubled-cell detector
that does not misidentify smooth extrema in the solution can be challenging.
This can lead to problems, such as a smearing out of the density maximum at the
center of a star.

In the context of an $hp$-adaptive DG method, however, the AMR criteria can
also be used to inform the troubled-cell detector. When the solution is not
smooth (i.e., the modal coefficients do not fall off rapidly enough), the AMR
algorithm will reduce the order $N$ of the element and trigger $h$-refinement.
High-order elements, then, have smooth solutions and do not require limiting.
In our manually refined grid, we apply the limiter only to elements with
$N \le 2$ in any spatial direction.

While our choices of grid setup and limiter application are not fully general,
they are representative of the outcome from a more general AMR DG code. Our
results are an exploration and will serve to inform the choices made in a
future AMR update to \textsc{SpECTRE} (the new DG code mentioned in Sec.\
\ref{sec:introduction}).

\section{Evolution of GR-hydro}
\label{sec:grhydro}

\subsection{Spacetime geometry}
\label{sec:st}

\subsubsection{Generalized harmonic equations}

We evolve the spacetime geometry using the generalized harmonic formulation of
Einstein's equations \cite{Friedrich1985,Pretorius2005c,Gundlach2005}. We use a
first-order representation of the system \cite{Lindblom2006} in which the
evolved variables are the spacetime metric $g_{\mu\nu}$, its spatial first
derivatives $\Phi_{i\mu\nu} = \partial_i g_{\mu\nu}$, and its first derivative
$\Pi_{\mu\nu} = - t^{\sigma} \partial_{\sigma} g_{\mu\nu}$ along the (timelike,
future-directed) normal $t^{\sigma}$ to the constant-$t$ hypersurface. The
complete equations for $\partial_t g_{\mu\nu}$, $\partial_t \Phi_{i\mu\nu}$,
and $\partial_t \Pi_{\mu\nu}$%
\footnote{Where we use $g_{\mu\nu}$, the cited papers use $\psi_{\mu\nu}$ to
denote the spacetime metric.}
in a vacuum spacetime can be found in Ref.\ \cite{Lindblom2006}; when coupling
the spacetime to matter, we add the source term
\begin{equation}
\partial_t \Pi_{\mu\nu} =
\begin{pmatrix}
\text{vacuum} \\
\text{terms}
\end{pmatrix}
- 2 \alpha \left(T_{\mu\nu}
  - \frac{1}{2} g_{\mu\nu} T^{\rho\sigma} g_{\rho\sigma}\right).
\end{equation}
The DG method for this system of equations takes the form \eqref{eq:dgfosh}.
The characteristic variables and speeds for the system, used in the upwind
numerical flux shown below, are also given in Ref.\ \cite{Lindblom2006}.

For the cases we present in this paper, the natural coordinates of the initial
data are well suited to prolonged time evolution. The generalized harmonic
gauge function $H_{\sigma}$, which specifies the coordinates, is therefore
independent of time. Its precise form will depend on the data being evolved.
The constraint-damping parameters $\gamma_0$ and $\gamma_2$, which constrain
the evolution of the coordinates and the growth of short-wavelength
perturbations, respectively, are also problem dependent. Following Ref.\
\cite{Lindblom2006}, we fix the parameter $\gamma_1$ to $-1$ because this makes
the generalized harmonic system linearly degenerate.

\subsubsection{Upwind flux}

As the solutions to the Einstein equations are smooth, we use an upwind
numerical flux, which sets the flux through the boundary according to the
propagation direction of each characteristic variable. The characteristic
decomposition of the system is given by
\begin{equation}
A^a n_a u = S \Lambda S^{-1} u,
\end{equation}
where $S$ diagonalizes the product $A^a n_a$%
\footnote{
At each point, we treat the background spacetime (i.e., $A^a n_a$) as constant
and compute the wave decomposition of the state vector $u$ by treating it as
a perturbation.};
i.e., the $i^{\text{th}}$ column of $S$ is the right eigenvector of $A^a n_a$,
with eigenvalue $\lambda_i$. Physically, the $S^{-1} u$ are the characteristic
variables of the system, and $\lambda_i$ are the associated propagation speeds
with respect to the normal $n_a$. The diagonal matrix
$\Lambda = \text{diag}(\lambda_1,...,\lambda_n)$ holds these eigenvalues and
can be separated by the sign of the eigenvalues,
$\Lambda = \Lambda^{+} + \Lambda^{-}$. At a boundary with two edge states
$u^{\text{L}}$ and $u^{\text{R}}$ and a normal $n_a$ directed toward the R
state, the upwind numerical flux takes the form
\begin{equation}
(A^a n_a u)^{\text{upwind}} = S \left( \Lambda^{+} S^{-1} u^{\text{L}}
  + \Lambda^{-} S^{-1} u^{\text{R}} \right),
\end{equation}
so that characteristic variables propagating left to right (in the direction of
$n_a$, with $\lambda_i > 0$) are set from the $u^{\text{L}}$ state, whereas
variables propagating right to left (with $\lambda_i < 0$) are set from
$u^{\text{R}}$.

\subsection{Hydrodynamics}
\label{sec:hydro}

\subsubsection{Relativistic fluid equations}

We treat the matter as a perfect fluid. Its stress-energy tensor takes the form
\begin{equation}
T_{\mu\nu} = \rho h u_{\mu} u_{\nu} + p g_{\mu\nu},
\end{equation}
where $\rho$ is the fluid's rest-frame mass density, $p$ is the pressure, and
$h = 1+\epsilon+p/\rho$ is the relativistic specific enthalpy, with $\epsilon$
the specific internal energy density. From the fluid's 4-velocity
$u^{\mu} = W(1, v^i)$, we define the lower 3-velocity components
$v_i = \gamma_{ij} v^j$ and the Lorentz factor
$W = \alpha u^0 = 1/\sqrt{1-v_iv^i}$. An equation of state (EOS) relates $p$,
$\rho$, and $\epsilon$; we use an ideal-gas EOS $p = (\Gamma - 1)\rho\epsilon$,
with $\Gamma$ the adiabatic index. In the absence of shocks, this is equivalent
to a polytropic EOS where $p = \kappa \rho^{\Gamma}$, with $\kappa$ some
constant.

The dynamics of the fluid are governed by the relativistic Euler equations. We
use the Val\`{e}ncia form of these equations \cite{lrr-2008-7}, with conserved
quantities $\{D, S_i, \tau\}$: the mass-energy density, momentum density, and
internal energy, as measured by a generalized Eulerian observer. These are
given by
\begin{equation}
\label{eq:releulerU}
\sqrt{\gamma}u
=
\begin{pmatrix}
\tilde{D} \\
\tilde{S}_i \\
\tilde{\tau}
\end{pmatrix}
=
\begin{pmatrix}
\sqrt{\gamma} W \rho \\
\sqrt{\gamma} W^2 \rho h v_i \\
\sqrt{\gamma} \left( W^2 \rho h - p - W\rho \right)
\end{pmatrix}.
\end{equation}
We follow the convention of using tildes to indicate ``densitized'' variables,
$\tilde{X} \equiv \sqrt{\gamma} X$. The corresponding flux vector and source
term are
\begin{align}
& \sqrt{\gamma}F^a
=
\begin{pmatrix}
\tilde{D} v^a_{\text{tr}} \\
\tilde{S}_i v^a_{\text{tr}} + \sqrt{\gamma} \alpha p \delta^a_i \\
\tilde{\tau} v^a_{\text{tr}} + \sqrt{\gamma} \alpha p v^a
\end{pmatrix}
\label{eq:releulerF}
\\
& \sqrt{\gamma}s
=
\begin{pmatrix}
0 \\
(\alpha/2) \tilde S^{lm} \partial_i \gamma_{lm} + \tilde S_k \partial_i \beta^k
  - \tilde E \partial_i \alpha \\
\alpha \tilde S^{lm}K_{lm} - \tilde S^l \partial_l \alpha
\end{pmatrix}.
\label{eq:releulerS}
\end{align}
Here, $v_{\text{tr}}^a = \alpha v^a - \beta^a = u^a / u^0$ is the transport
velocity relative to the coordinates; $S^{lm}$ and $E$ are components of the
stress energy,
\begin{align}
\tilde{S}^{lm} &= \sqrt{\gamma} T^{lm} = \sqrt{\gamma} \rho h W^2 v^l v^m
  + \sqrt{\gamma} p \gamma^{lm}
\\
\tilde{E} &= \sqrt{\gamma} n^{\mu} n^{\nu} T_{\mu\nu}
  = \sqrt{\gamma} \rho h W^2 - \sqrt{\gamma} p;
\end{align}
and $K_{lm}$ is the usual extrinsic curvature of the constant-$t$ hypersurface.
The system of equations is evolved according to the discretized form
\eqref{eq:dgfinal}, with the densitized conserved variables
$\{\tilde{D}, \tilde{S}_i, \tilde{\tau}\}$ serving as the primary variables in
the code. The characteristic speeds, used in the numerical fluxes shown below,
are given in Ref.\ \cite{Banyuls1997}.

Solving for the primitive variables $\{\rho,v_i,\epsilon\}$ from
$\{D,S_i,\tau\}$ requires root finding and may additionally require
``atmosphere fixing'' in regions of low density where the inversion may be
numerically poorly behaved. We follow the inversion and fixing procedure of
Ref.\ \cite{Galeazzi:2013mia}, Appendix C. This fixing procedure takes grid
points where the low-density state $\{D,S_i,\tau\}$ does not correspond to a
physical state $\{\rho,v_i,\epsilon\}$ and alters the conserved variables to
recover a physical state. Additionally, a small (i.e., dynamically negligible)
floor $\rho_{\text{atmo}}$ is set on the fluid density, ensuring that round-off
level errors are controlled. In the test problems of Sec.\ \ref{sec:tests},
fixing is not needed; we set $\rho_{\text{atmo}}$ to 0. For the NS evolutions
of Sec.\ \ref{sec:results}, fixing is necessary outside the star; we give the
parameters of the fixing within that section.

\subsubsection{Numerical fluxes}

For the fluid, we use a numerical flux chosen to approximately solve the
Riemann problem corresponding to the discontinuity between elements. As before,
we label the two states at the boundary as $u^{\text{L}}$ and $u^{\text{R}}$,
and the normal $n_a$ points toward the R state. A popular choice of numerical
flux, because of its robustness and simplicity, is the local Lax-Friedrichs
(LLF) flux. This flux is computed according to
\begin{equation}
(F^{a*} n_a)^{\text{LLF}}
  = \frac{F^a(u^{\text{L}}) n_a + F^a(u^{\text{R}}) n_a}{2}
    - \frac{C}{2} \left( u^{\text{R}} - u^{\text{L}} \right),
\end{equation}
where $C = \max (|\lambda_i(u^{\text{L}})|, |\lambda_i(u^{\text{R}})|)$ is the
largest speed across the interface. The speeds $\lambda_i$ are again the
eigenvalues of the flux Jacobian (see the upwind flux discussion above, with
$A^a \to \partial F^a / \partial u$). We maximize over the $\lambda_i$ on both
sides of the interface, but independently at each interface grid point.

A more sophisticated numerical flux, which includes an approximate treatment of
the system's underlying wave structure, is given by Harten, Lax, and van Leer
(HLL) \cite{HLL,toro2013riemann},
\begin{align}
\nonumber
(F^{a*} n_a)^{\text{HLL}} = &\frac{c_{\text{max}} F^a(u^{\text{L}}) n_a
  + c_{\text{min}} F^a(u^{\text{R}}) n_a}{c_{\text{max}} - c_{\text{min}}}
\\
&- \frac{c_{\text{max}} c_{\text{min}}}{c_{\text{max}}
  - c_{\text{min}}} \left( u^{\text{R}} - u^{\text{L}} \right).
\end{align}
Here, $c_{\text{min}}$ and $c_{\text{max}}$ are estimates for the fastest left-
and right-moving signal speeds, respectively. We use the simple estimates
\cite{davis1988simplified}, computed pointwise,
\begin{align}
\nonumber
c_{\text{min}} &= \min \left( \lambda_i(u^{\text{L}}), \lambda_i(u^{\text{R}}),
  0 \right)
\\
c_{\text{max}} &= \max \left( \lambda_i(u^{\text{L}}), \lambda_i(u^{\text{R}}),
  0 \right).
\end{align}
Note that the HLL flux reduces to upwinding when all $\lambda_i$ share the same
sign, i.e., all characteristic variables are propagating in the same direction.

We find that the LLF and HLL fluxes give very similar results in most of the
cases we tested (for an exception, see the supersonic accretion flow test in
Sec.\ \ref{sec:hydrotests}) and conclude that the use of an approximate
solution to the Riemann problem does not introduce a significant error in these
problems. The results presented in this paper are computed using the HLL flux.

\subsubsection{Limiters}
\label{sec:limiters}

In this work, we use and compare two limiters. The first is the simple, but
also low-order, $\Lambda \Pi^1$ slope limiter \cite{Hesthaven2008, Cock01},
which we will refer to simply as minmod. This limiter computes several
estimates for the slope of the solution on each element, and then, in elements
where these estimates indicate the presence of oscillations in the solution, it
acts to reduce the slope. Taking the 1D case as example, we write the solution
$u^k$ on the $k^{\text{th}}$ element as a series expansion,
\begin{equation}
u^k = \bar{u}^k + u_1 (x-x_0) + \mathcal{O}(x-x_0)^2,
\end{equation}
where $\bar{u}^k$ is the element-averaged mean of $u^k$, $u_1$ is the mean
slope, and $x_0$ is the center of the element. The minmod limiter's slope
estimates are
\begin{equation}
a_1 = u_1
, \quad
a_2 = \frac{\bar{u}^{k+1} - \bar{u}^k}{h / 2}
, \quad
a_3 = \frac{\bar{u}^k - \bar{u}^{k-1}}{h / 2},
\end{equation}
where $h$ is the width of the element. The limiter selects the estimate with
the smallest absolute value (or 0, if the three estimates differ in sign). If
the selected estimate is not the original slope $u_1$, the limiter activates by
reducing the slope $u_1$ to the selected estimate (or 0) and discarding any
higher-order terms in the approximation. On elements with order $N > 1$, we use
the $\Lambda \Pi^N$ generalization of the limiter described in Ref.\
\cite{Hesthaven2008}. We do not use the ``total variation bound''
generalization, which sets a scale below which oscillations are tolerated,
since we find that it is not robust at star surfaces.

In higher dimensions, the 1D limiter is applied to each direction in turn.
After this process, the limited solution may occasionally correspond to a
nonphysical state. When this occurs, we further reduce the slope until the
following are satisfied throughout the element:
$\text{min}(D) > \rho_{\text{atmo}}$, $\text{min}(\tau) > 0$, and
$S^2 < \tau(\tau + 2D)$.

For evolutions on deformed grids, we apply the 1D limiter along each direction
of the reference $\bar{\textbf{x}}$ coordinates. This choice of coordinates
leads to a straightforward computation of the minmod slope estimates, because
the series representation of $u^k$ takes a simple form, and the element's upper
and lower neighbors in each direction are well defined. However, the choice
introduces a (relatively small) violation of conservation: the limiter will
conserve the means $\bar{u}^k$ with respect to the reference $\bar{\textbf{x}}$
coordinates, but the means with respect to the ``global'' $\textbf{x}$
coordinates will in general be modified after the limiter activates. This can
be understood by noting that the means in the two coordinate systems are
differently sensitive to the shape of the function $u^k$:
\begin{equation}
\bar{u}^k|_{\bar{\textbf{x}}} = \frac{\int u^k d^3\bar{x}}{\int d^3\bar{x}}
\quad
\text{vs}
\quad
\bar{u}^k|_{\textbf{x}}
  = \frac{\int u^k d^3x}{\int d^3x}
  = \frac{\int u^k J d^3\bar{x}}{\int J d^3\bar{x}}.
\end{equation}
We explored two simple corrections to the minmod limiter that restore
conservation in the $\textbf{x}$ coordinates. The first correction limits the
Jacobian-weighted solution $Ju^k$ instead of $u^k$; the second shifts the
postlimiting solution $u^k \to u^k + \delta u^k$, with $\delta u^k$ a constant
computed to restore the prelimiting mean $\bar{u}^k$. Both of these corrections
successfully restore the limiter's conservative properties, but we found that
they also introduced long-timescale instabilities at the surface of the star
--- we note that Radice and Rezzolla \cite{Radice:2011qr} also found poor
behavior when using similar corrections with the simple minmod limiter.
Consequently, we do not use these corrections in our simulations. Instead, we
will quantify the error in maintaining conservation when presenting our
results.

The second limiter we consider is that of Moe \emph{et al}.\
\cite{moe2015simple}, henceforth MRS. This limiter acts by scaling the
conserved variables $u$ about their means $\bar{u}$,
\begin{equation}
\label{eq:mrs}
u \to \bar{u} + \theta (u - \bar{u}),
\end{equation}
with $\theta \in [0,1]$ determined from analysis of the minima and maxima of
the solution in the immediate neighborhood of the element. A tolerance function
$\alpha(h)$ sets the scale below which oscillations are tolerated; we use the
function $\alpha(h) = 100 h^{3/2}$ for the cases presented in this paper, as it
performs well on many different test problems.

We obtain best results when computing $\theta$ from the primitive variables, as
MRS recommend. However, care must be taken when computing the primitive
variables, as the fluid state may be unphysical until limited. We ``prelimit''
by applying an additional scaling of the form \eqref{eq:mrs} to the conserved
variables. The steps below restore a physical state and ensure the inversion
procedure is well posed:
\begin{enumerate}
\item If $\min(D) < \rho_{\text{atmo}}$ or $\min(\tau) < 0$,
  scale to fix these violations.
\item If $S_i S^i > \tau ( \tau + 2 D )$ at any grid point, scale to fix this
  violation. This requires solving a quadratic equation for $\theta$.
\item If the inversion to primitive variables encounters any of the errors
  outlined in Ref.\ \cite{Galeazzi:2013mia}, Appendix C (this is rare), scale
  again with $\theta = 1/2$.
\end{enumerate}
This procedure is conservative by construction, and we find it to be robust.
After this prelimiting step, we compute the primitive variables and limit
according to the MRS prescription. We handle deformed grids as for the minmod
limiter, by computing the means in the reference $\bar{\textbf{x}}$ coordinates
and incurring some error due to loss of conservation. As with minmod, attempts
to reformulate the limiter to restore conservation (we tried the simple
approach of computing the MRS means directly with respect to the $\textbf{x}$
coordinates as well as the same two reformulations described for minmod) were
not stable at the star surface.

We apply the limiter to the fluid variables at the end of each time-stepper
substep. As described in Sec.\ \ref{sec:philosophylimiter}, we may not apply
the limiter to every element, choosing instead to mimic an AMR scheme in which
high-order elements are known to be smooth. The use of more complex,
higher-order, limiters, e.g., subcell methods
\cite{Radice:2011qr,Bugner:2015gqa} or the compact-stencil WENO
\cite{Zhong2013:WENODG} and HWENO \cite{Zhu2016:HWENODG} limiters, will be the
subject of future investigation.

\subsection{Combined GR-hydro system}

For self-consistent NS evolutions, the equations of the generalized harmonic
and relativistic Euler systems are each treated as described above and are
evolved in parallel. The two systems couple via their respective source terms
and the geometry terms in the hydrodynamics flux $F^a(u)$. We compute the
characteristic speeds independently for each system, leaving out the
cross-coupling arising from the off-diagonal
$\partial F^a_{\text{hydro}} / \partial u_{\text{GR}}$ flux Jacobian terms.
When the fluid variables require limiting, the limiter is applied to the fluid
variables only, and the spacetime variables are left unmodified.

\subsection{Filtering}

The use of inexact quadratures to obtain an efficient DG scheme may result in
numerical instabilities caused by aliasing. Where these numerical instabilities
exist, we address them by filtering the higher modes in the solution's modal
representation. We use an exponential filter, e.g., in one dimension,
\begin{equation}
\label{eq:filter}
\hat{u}_i \to F(i) \hat{u}_i = \exp{(-\alpha (i/N)^s)} \hat{u}_i,
\end{equation}
where $\alpha$ controls the strength of the filter's effect and $s$ is an even
integer controlling how many modes are affected. In $d>1$ dimensions, we take
advantage of the tensor-product basis to apply the filter dimension by
dimension; this gives $d$ exponentials. On deformed grids, we filter the
Jacobian-weighted solution $Ju$ and then divide by $J$, so that the operation
remains conservative. We apply the filter at the end of each complete time step
to the components of $u$ and on the elements that show numerical instability.

\subsection{Time stepping}

We use the third-order strong stability-preserving Runge-Kutta scheme of Ref.\
\cite{gottlieb2001strong} for the time integration. Given the solution $u^n$ at
time $t^n$, the solution $u^{n+1}$ at time $t^{n+1} = t^n + \Delta t$ is
computed as
\begin{align}
\nonumber
u^{(1)} &= u^n + \Delta t L(u^n)
\\
\nonumber
u^{(2)} &= \frac{3}{4} u^n + \frac{1}{4}\left[u^{(1)}
  + \Delta t L(u^{(1)})\right]
\\
u^{n+1} &= \frac{1}{3} u^n + \frac{2}{3}\left[u^{(2)}
  + \Delta t L(u^{(2)})\right].
\end{align}
Here, $L(u) = du/dt$ is computed from expressions \eqref{eq:dgfinal} for the
fluid variables or \eqref{eq:dgfosh} for the spacetime variables.

In all cases presented, the initial $t=0$ data are computed by pointwise
evaluation of a known state. The limiter is applied to the initial data and at
the end of every subsequent substep. Filtering is done at the end of full time
steps.

\section{Code tests}
\label{sec:tests}

In this section, we present a selection of benchmark tests that we use to
validate our implementation of the DG method within \textsc{SpEC}.

We first show tests of vacuum spacetime evolution. From a family of gauge wave
evolutions at varying resolutions, we verify that the method converges to the
exact solution at the expected rate. Next, by evolving a Kerr (i.e., isolated
and spinning) BH over long timescales, we show the stability of the algorithm.

We then show our tests of the hydrodynamics implementation. We again verify the
convergence rate of the errors, now with a generalized Bondi problem in which
the fluid undergoes spherically symmetric accretion onto a Schwarzschild BH.
This test verifies the fluid equations as well as the sourcing of the fluid by
the spacetime curvature. We then show standard shock tests in one and two
dimensions, comparing the effectiveness of the implemented limiters.

In these tests, whenever possible, we compare the numerical solution to an
exact solution, and we use their difference as an error measure. We report a
normalized error $\text{err}[X]$ in a quantity $X$, defined as
\begin{equation}
\text{err}[X] = \norm{X - X_{\text{exact}}} \big{/} \norm{X_{\text{exact}}}.
\end{equation}
Here, $\norm{X}$ is the $L^2$-norm, evaluated pointwise by direct summation over
every node of the computational grid,
\begin{equation}
\norm{X}^2 = \frac{1}{N_{\text{nodes}}} \sum_{i=0}^{N_{\text{nodes}}} X_i^2.
\end{equation}
When $X$ is a vector or tensor quantity, we compute a component-wise norm
$\norm{X}^2 = \norm{X_0}^2 + \norm{X_1}^2 + ...$, rather than the physical
norm $X_a X^a$. When $X_{\text{exact}} = 0$ so that we cannot define the
normalized error, we instead use $\norm{X}$ as our error measure.

\subsection{Spacetime tests}
\label{sec:grtests}

\subsubsection{Gauge-wave test}

The spacetime of the ``apples to apples'' gauge-wave test
\cite{Alcubierre:2003pc}, obtained via a nonlinear, plane-wave transformation
of Minkowski space, takes the form
\begin{equation}
ds^2 = -(1+a)dt^2 + (1+a)dx^2 + dy^2 + dz^2,
\end{equation}
with
\begin{equation}
a = A \sin[2\pi(x-t)].
\end{equation}
We show results for a wave of amplitude $A=0.1$ on a unit-cube domain with
extents $[0,1]^3$. As the gauge wave is harmonic, the generalized harmonic
gauge function $H_{\sigma}$ is zero. We set the constraint-damping parameters
$(\gamma_0, \gamma_1, \gamma_2)$ to $(1, -1, 1)$, values that give stable
evolutions over long timescales (up to at least $t_{\text{fin}}=1000$, or 1000
crossing times). For the convergence study, however, we measure the error in
the spacetime metric $g_{\mu\nu}$ at a final time $t_{\text{fin}}=10$, after
evolution with time steps of size $\Delta t = 10^{-4}$. This time step
corresponds to $\Delta t / \Delta x_{\text{min}} \simeq 0.074$ for the
highest-resolution case in the convergence study ($K=128$, $N=4$).

\begin{figure}[tb]
\centering
\includegraphics[width=\columnwidth]{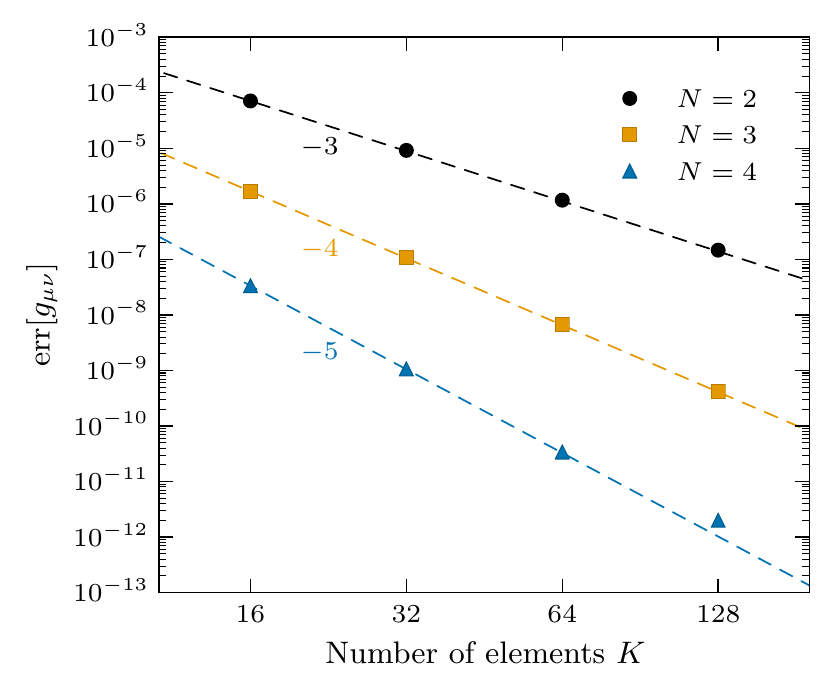}
\caption[The error in $g_{\mu\nu}$ as a function of the number of elements
  ($h$-refinement) for the gauge-wave test of Einstein's equations.]{
  The error in $g_{\mu\nu}$ as a function of the number of elements
  ($h$-refinement) for the gauge wave test of Einstein's equations. The symbols
  indicate the measured error norms for methods of order $N$ = 2, 3, 4. The
  dashed lines, normalized to the $K=16$ data, indicate the expected error
  scaling for third-, fourth-, and fifth-order convergence.
}
\label{fig:gwhconv}
\end{figure}

We show in Fig.\ \ref{fig:gwhconv} the convergence under $h$-refinement,
measured for elements of order $N=2,3,4$. For a base resolution, we partition
the unit-cube domain into $16$ elements along the $x$ direction; we $h$ refine
by further splitting each element along $x$, reducing the element's width $h$
in half each time. We do not split in $y$ or $z$ --- the anisotropic refinement
is chosen to match the $x$-only dependence in the problem. For each order $N$
of the DG method, we compare our measurements to the theoretical scaling of the
error (see, for instance, Ref.\ \cite{Hesthaven2008}),
\begin{equation}
\label{eq:convergencerelation}
\text{err}[g_{\mu\nu}] \leq C h^{N+1} \propto 1/K^{N+1},
\end{equation}
for some constant $C$. We find excellent agreement between the measured and
expected convergence rates. The highest-resolution case ($K=128$ and $N=4$) has
a slightly larger error, having reached the round-off level error in the
derivatives of the spacetime.

\begin{figure}[tb]
\centering
\includegraphics[width=\columnwidth]{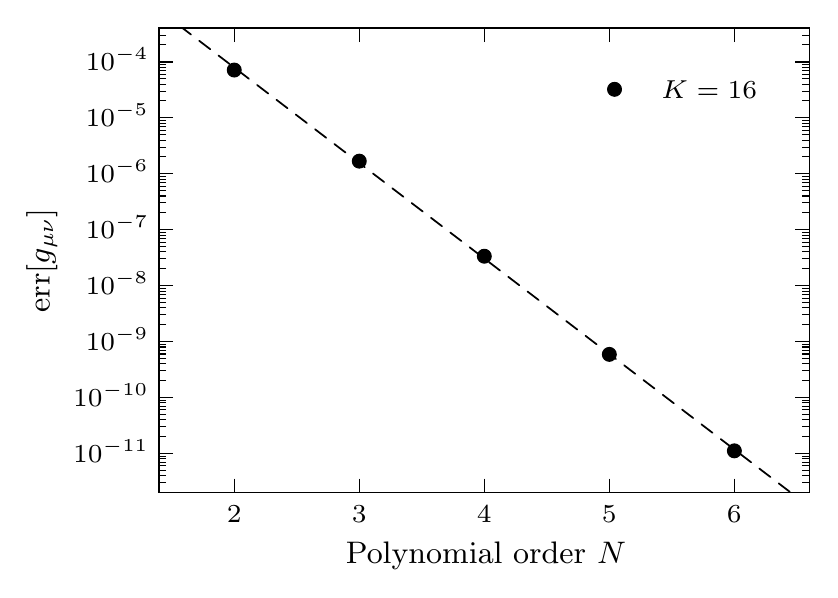}
\caption[The error in $g_{\mu\nu}$ as a function of the order of approximation
  ($p$-refinement) for the gauge-wave test of Einstein's equations.]{
  The error in $g_{\mu\nu}$ as a function of the order of approximation
  ($p$-refinement) for the gauge wave test of Einstein's equations. The number
  of elements is fixed at $K=16$. The dots indicate the measured errors; the
  dashed line is a fit demonstrating the exponential decrease in the error with
  $N$.
}
\label{fig:gwpconv}
\end{figure}

In Fig.\ \ref{fig:gwpconv}, we show the convergence under $p$-refinement,
obtained by increasing the order $N$ of the DG method while maintaining the
base resolution of 16 elements. We expect the errors to decrease exponentially
with the order $N$ and recover this behavior in our measurements. This result
demonstrates the spectral convergence of the DG method for smooth solutions.

\subsubsection{Kerr black hole}

We next evolve the spacetime of a Kerr BH, described by the Kerr metric in
Kerr-Schild coordinates \cite{MTW}. The BH has spin
$\vec{a}=(0.1,0.2,0.3) M_{\text{BH}}$ with magnitude
$a \approx 0.374 M_{\text{BH}}$, not aligned with any grid symmetries. We use
units where $M_{\text{BH}}=1$.

The domain is a hollow spherical shell that excises the singularity within the
BH. In terms of the coordinate radius $r$, the domain extends from
$r_{\text{in}} = 1.8$ (just inside the event horizon) to $r_{\text{out}} = 32$.
At the inner boundary, all the characteristics of the system are outgoing
(i.e., leaving the domain, toward the singularity), so no boundary condition
needs to be imposed. Physically, no information enters the simulation from the
interior of the BH. At the outer boundary, we impose the analytic solution as a
Dirichlet boundary condition.%
\footnote{
We do not use the constraint-preserving boundary conditions typically used in
\textsc{SpEC} simulations, because these are boundary conditions on
$\partial_t u$, rather than $u$, and so would require a modification of the DG
formulation.}
We choose constraint damping parameters
\begin{align}
\gamma_0 &= 3 \exp[-(r/8)^2/2] + 0.1
\\
\gamma_1 &= -1
\\
\gamma_2 &= \exp[-(r/8)^2/2] + 0.1.
\end{align}
The generalized harmonic gauge function
$H_{\sigma} = \Gamma_{\sigma} \equiv g^{\mu\nu} \Gamma_{\sigma \mu\nu}$ is the
trace of the Christoffel symbols of the Kerr-Schild metric; it is constant in
time.

\begin{figure}[tb]
\hspace*{\fill}
\subfloat[Projection]{
\includegraphics[width=0.45\columnwidth]{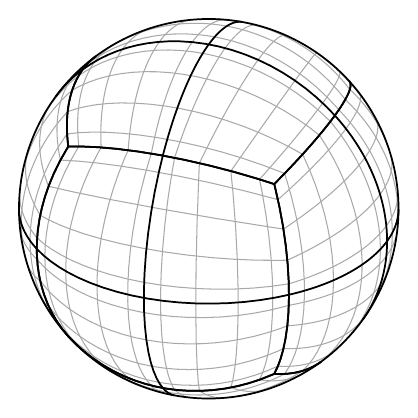}
}
\hfill
\subfloat[Equatorial cut]{
\includegraphics[width=0.45\columnwidth]{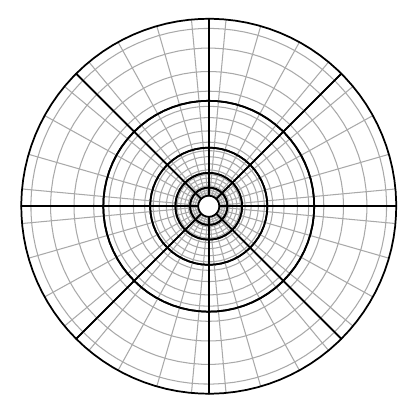}
}
\hspace*{\fill}
\caption[The grid structure for the Kerr BH evolution test.]{
  The grid structure for the Kerr BH evolution test. Shown are (a) a projected
  view and (b) an equatorial cut. The black lines show the element boundaries,
  and the light grey lines show the Gauss-Legendre-Lobatto grid within each
  element for order $N=5$.
}
\label{fig:bhdomains}
\end{figure}

We set up a cubed-sphere grid on this domain, using the mappings from Appendix
\ref{app:cubedsphere}. The wedges of the cubed sphere are split radially into
five concentric shells located between the surfaces $r$ = 1.8, 3.2, 5.7, 10,
18, 32, and then tangentially into $2\times2$ angular portions, for a total of
120 elements. The tangential coordinates of each wedge are mapped to obtain an
equiangular grid, as this is a more optimal distribution for the grid points on
the spherical surface. We show in Fig.\ \ref{fig:bhdomains} two views of this
grid: on the left a projected view showing the angular structure on a
constant-radius surface, and on the right an equatorial cut showing the radial
structure. The increasing density of grid points toward the center of the
domain helps to resolve the stronger spacetime curvature near the BH.

\begin{figure}[tb]
\centering
\includegraphics[width=\columnwidth]{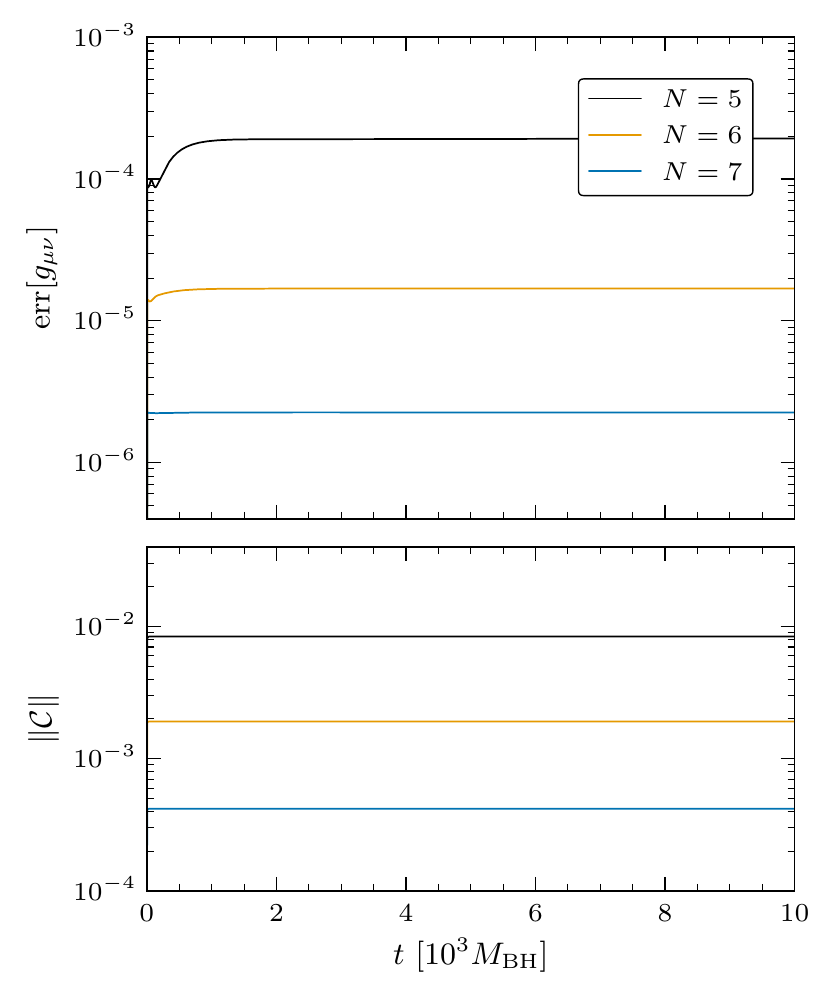}
\caption[The errors during the Kerr BH evolution test.]{
  The errors during the Kerr BH evolution test. The top panel shows the error
  in the spacetime metric $g_{\mu\nu}$ for three different orders of the DG
  method. The lower panel shows the dimensionless norm of the generalized
  harmonic energy constraint at the same three orders.
}
\label{fig:bhstability}
\end{figure}

In Fig.\ \ref{fig:bhstability}, we show the stability of the Kerr BH evolution
by monitoring the simulation errors over a duration of $10^4 M_{\text{BH}}$.
We carry out the simulation using elements of order $N$ = 5, 6, and 7; the
time-step size is $\Delta t = 10^{-2}$, giving
$\Delta t / \Delta x_{\text{min}} \simeq 0.15$ for the $N=7$ case. The figure's
top panel shows the error $\text{err}[g_{\mu\nu}]$ in the spacetime metric, a
measure of the solution's drift from the exact value. The bottom panel shows
the dimensionless norm $\norm{\mathcal{C}}$ of the generalized harmonic energy
constraint \cite{Lindblom2006}, a measure of how well the numerical solution at
each constant-$t$ slice satisfies Einstein's equations. After a rapid settling
of the solution to its numerical equilibrium, we see clear convergence in the
error quantities. We conclude that the method is convergent and stable up to at
least $t=10^4 M_{\text{BH}}$ and, we presume, forever.

\subsection{Relativistic hydrodynamics tests}
\label{sec:hydrotests}

\subsubsection{Spherical accretion onto black hole}

In the relativistic Bondi problem, an ideal gas accretes radially onto a
nonrotating BH. The feedback from the fluid onto the spacetime is ignored: the
BH mass is constant, and the spacetime is Schwarzschild. We use Kerr-Schild
coordinates, and again we set $M_{\text{BH}}=1$. The analytic profile for the
fluid flow is presented by Michel \cite{Michel:1972}; following Ref.\
\cite{Banyuls1997}, we pick a solution for a $\Gamma = 5/3$ ideal gas with the
sonic point and mass accretion rate given by $r_{\text{crit}} = 200$ and
$\dot{M} = 10^{-3}$. We measure the error in the conserved relativistic
density $\tilde{D}$ at a final time $t_{\text{fin}}=100$, after evolution
with time steps of size $\Delta t = 5\times10^{-3}$. This time step corresponds
to $\Delta t / \Delta x_{\text{min}} \simeq 0.15$ for the highest-resolution
case in the convergence study ($K=120\times4^3$, $N=4$).

We evolve the fluid in a hollow spherical shell extending from
$r_{\text{in}} = 1.8$ (just inside the event horizon), to
$r_{\text{out}} = 12$. The sonic point in the accretion flow is located
outside this region, so the flow is smooth and supersonic throughout the
simulation domain. In this test problem, we obtain significantly more accurate
results when using the HLL numerical flux (vs LLF), as the supersonic flow is
best represented by the HLL upwinding limit. At the inner boundary, the
characteristics of the fluid system are outgoing (i.e., leaving the domain into
the BH), so no boundary condition needs to be applied. At the outer boundary,
we impose the analytic solution as a boundary condition.

We use a cubed-sphere grid similar to that of the Kerr BH test above. At the
base resolution, we divide the domain into five spherical shells between the
surfaces located at radii $r$ = 1.8, 2.7, 4, 6, 9, 12, and we split each
wedge into $2\times2$ angular portions. The tangential coordinates are again
mapped to obtain an equiangular grid.

\begin{figure}[tb]
\centering
\includegraphics[width=\columnwidth]{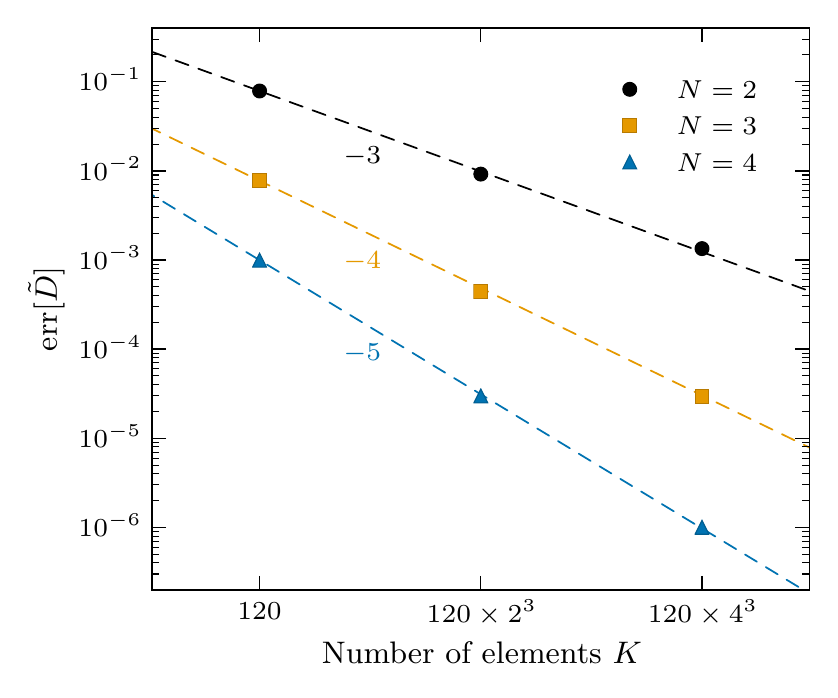}
\caption[The error in the conserved density $\tilde{D}$ as a function of the
  number of elements ($h$-refinement) for the spherical accretion test.]{
  The error in the conserved density $\tilde{D}$ as a function of the number of
  elements ($h$-refinement) for the spherical accretion test. The symbols
  indicate the measured error norms for methods of order $N$ = 2, 3, 4.
  The dashed lines, normalized to the $K = 120$ data, indicate the expected
  error scaling for third-, fourth-, and fifth-order convergence.
}
\label{fig:accrhconv}
\end{figure}

We show in Fig.\ \ref{fig:accrhconv} the convergence under $h$-refinement of
this grid, for elements of order $N$ = 2, 3, 4. We $h$ refine by splitting each
element into $2^3$ smaller elements; we split geometrically in radius according
to $r_{\text{split}} = \sqrt{r_{\text{lower}} r_{\text{upper}}}$ and linearly
in the tangential directions. As the elements are not uniform, this choice of
radial split is not unique, but we find it gives reduced error compared to a
linear split $r_{\text{split}} = (r_{\text{lower}}+r_{\text{upper}})/2$. We
again see the errors converging at the expected rate.

\begin{figure}[tb]
\centering
\includegraphics[width=\columnwidth]{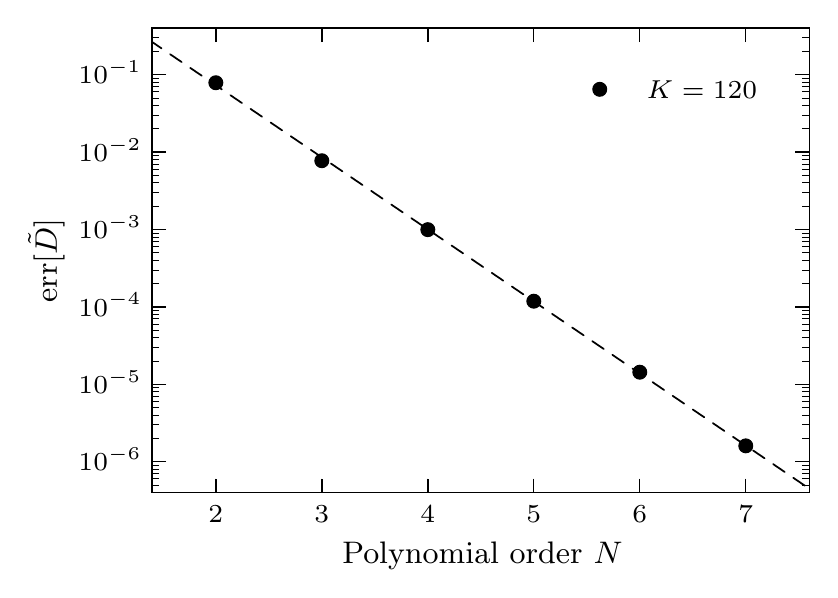}
\caption[The error in the conserved density $\tilde{D}$ as a function of the
  order of approximation ($p$-refinement) for the spherical accretion test.]{
  The error in the conserved density $\tilde{D}$ as a function of the order of
  approximation ($p$-refinement) for the spherical accretion test. The number
  of elements is fixed at $K = 120$. The dots indicate the measured errors; the
  dashed line is a fit demonstrating the exponential decrease in the error with
  $N$.
}
\label{fig:accrpconv}
\end{figure}

In Fig.\ \ref{fig:accrpconv}, we show the convergence under $p$-refinement.
Again, we use the base configuration of elements and increase the order $N$ of
the method from 2 to 7. We confirm that for this smooth fluid evolution
problem, the errors decrease exponentially with the order of the method.

\subsubsection{1D shock tube test}

We perform a standard 1D relativistic shock test problem, in which a
high-density and -pressure fluid expands into a low-density and -pressure
fluid. Following Ref.\ \cite{Duez:2008rb}, we take a $\Gamma = 5/3$ ideal gas
initially split at $x=0.5$ into left and right states characterized by
\begin{equation}
(\rho, v_x, p) =
\begin{cases}
(10,0,40/3), & x<0.5
\\
(1,0,0), & x>0.5
\end{cases}.
\end{equation}
The simulation domain is an interval $x \in [0,1]$, which we divide into
$K=160$ elements of order $N=2$. We evolve the shock until a final time
$t_{\text{fin}}=0.4$, with time steps
$\Delta t = 4\times10^{-3}$ ($\Delta t / \Delta x_{\text{min}} = 0.128$).

\begin{figure}[tb]
\centering
\includegraphics[width=\columnwidth]{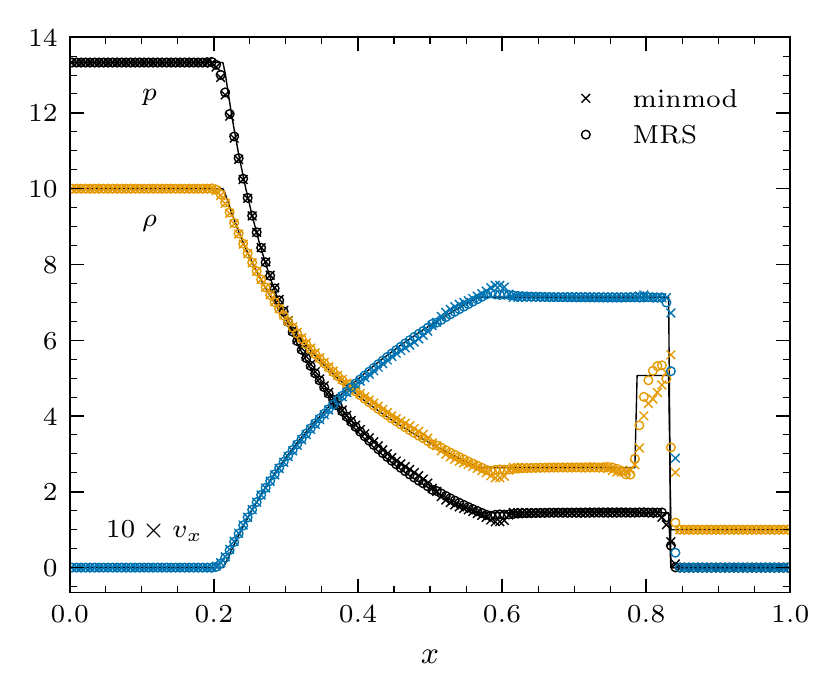}
\caption[Snapshot of the fluid variables in the shock tube test.]{
  Snapshot of the fluid variables in the shock tube test. The fluid pressure
  $p$, the rest-mass density $\rho$, and the velocity $v_x$ (scaled up
  $10\times$) are plotted after evolutions using the minmod and MRS limiters.
  The mean value on each element is shown. The exact solution to the problem is
  given by Centrella and Wilson \cite{Centrella1984} and is plotted here in the
  solid line.
}
\label{fig:sodshock}
\end{figure}

In Fig.\ \ref{fig:sodshock}, we show the profiles of $\rho$, $v_x$, and $p$ at
the final state, comparing the minmod and MRS limiters. Both limiters capture
the features of the shock profile. The minmod limiter produces a larger
overshoot at the main shock front and increased oscillation at the front end of
the rarefaction fan, a known behavior when applying this limiter to the
conserved variables (rather than characteristic variables
\cite{1989JCoPh..84...90C}).

\subsubsection{2D Riemann shock interaction test}

We next study a standard 2D Riemann problem in which two shocks and two contact
discontinuities interact. As in the 1D shock test, the fluid is a
$\Gamma = 5/3$ ideal gas. The initial conditions for the problem were first
generalized from Newtonian to relativistic hydrodynamics by Del Zanna and
Bucciantini \cite{DelZannaBucciantini2002a} and later modified by Mignone and
Bodo \cite{Mignone2005} to give a cleaner wave structure. The initial condition
divides the computational domain $[-1,1]^2$ into four quadrants, each of which
holds a constant fluid state,
\begin{equation}
(\rho, v_x, v_y, p) =
\begin{cases}
(0.5, 0, 0, 1), & x<0, y<0
\\
(0.1, 0, 0.99, 1), & x>0, y<0
\\
(0.1, 0.99, 0, 1), & x<0, y>0
\\
(\rho_1, 0, 0, p_1), & x>0, y>0
\end{cases},
\end{equation}
where the low-density state in the upper-right quadrant is defined by
$\rho_1 = 5.477875 \times 10^{-3}$ and $p_1 = 2.762987 \times 10^{-3}$.
We partition the domain into $200\times200$ elements of order $N=2$, and we
evolve until a final time $t_{\text{fin}}=0.8$ with time steps
$\Delta t = 10^{-3}$ ($\Delta t / \Delta x_{\text{min}} = 0.2$).

\begin{figure}[tb]
\centering
\includegraphics[width=\columnwidth]{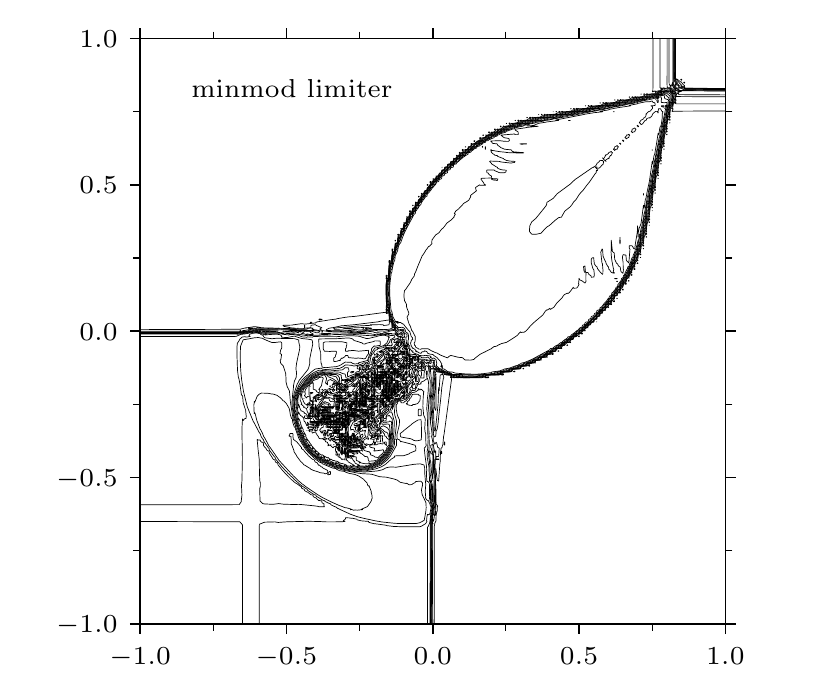}
\includegraphics[width=\columnwidth]{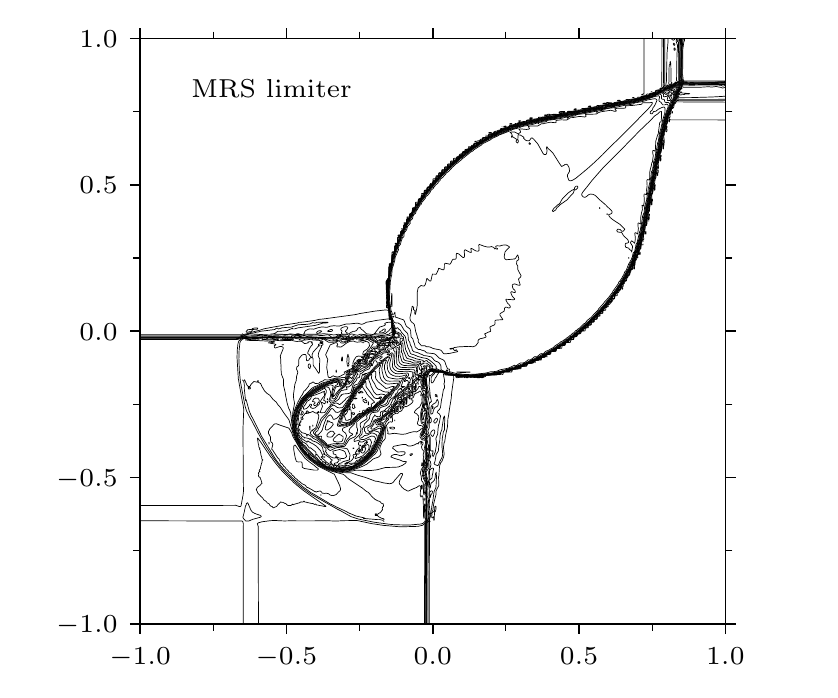}
\caption[The density $\rho$ in the 2D Riemann problem.]{
  The density $\rho$ in the 2D Riemann problem. The top panel is computed with
  the minmod limiter, and the bottom panel is computed with the MRS limiter.
  The plots each show 30 contour lines, equally spaced in $\log\rho$.
}
\label{fig:riemann}
\end{figure}

In Fig.\ \ref{fig:riemann}, we show contour plots of the density $\rho$ at the
final state. We interpolate the evolved $\rho$ onto a high-resolution uniform
grid on which the contours are computed. The results in the top panel are
computed with a minmod limiter, and those in the bottom panel are computed with
MRS. We find, qualitatively, excellent agreement between the results from the
two limiters; only the jet feature (in the lower-left quadrant) shows a clear
difference in resolution, with the MRS limiter producing a cleaner structure.

\section{Neutron star evolutions}
\label{sec:results}

Having verified the convergence and shock-capturing properties of our code, we
now present our main results: evolutions of an isolated, spherical NS using the
DG method. We first evolve the NS under the Cowling approximation, i.e.,
keeping the background spacetime fixed to the Tolman-Oppenheimer-Volkoff (TOV)
solution. This remains a challenging test of the hydrodynamics code's ability
to handle the discontinuity at the stellar surface. We then evolve the NS
self-consistently using the coupled GR-hydro system.

The initial data for the NS fluid and spacetime are found by integrating the
TOV equations
\cite{Tolman:1939jz,Oppenheimer:1939ne,MTW} for the mass-energy density
$\rho_E(R) \equiv \rho(R)(1+\epsilon(R))$, enclosed ADM mass $m(R)$, and metric
potential $\phi(R)$ in terms of the areal radius $R$. The spacetime metric is
given by
\begin{equation}
ds^2 = - e^{2\phi} dt^2 + \left(1 - \frac{2m}{R}\right)^{-1} dR^2
  + R^2 d\Omega^2.
\end{equation}
In computing the TOV solution, we describe the NS matter by a polytropic EOS.
When time evolving the solution, we return to the corresponding ideal-gas EOS.

The results presented throughout this section are for a star with
$\kappa = 100$ and $\Gamma=2$. The star has central mass density
$\rho_c = 1.28\times10^{-3}$, giving a stable, nonrotating TOV solution with
ADM mass $M_{\text{NS}} \simeq 1.4 M_{\odot}$ and areal radius
$R_{\text{NS}} \simeq 9.6 M_{\odot} \simeq 14$ km. Its radius in the isotropic
coordinates used during evolution is $r_{\text{NS}} \simeq 8.125 M_{\odot}$. In
this section, we use units where $M_{\odot}=1$.

For the NS evolutions, we use the atmosphere fixing from Ref.\
\cite{Galeazzi:2013mia}, Appendix C. We set the density cutoff
$\rho_{\text{atmo}}^{\text{cut}} = 10^{-15}$ so as to resolve 12 orders of
magnitude in density. Where the density falls below this cutoff, we set the
fluid to the ``atmosphere'' state where
$\rho = \rho_{\text{atmo}} = 10^{-16}$, $v_i = 0$, and $\epsilon = 0$.
Elsewhere, we constrain the specific internal energy to the range
$\kappa \rho \le \epsilon \le 100 \kappa \rho$, with $\kappa$ from the
polytrope describing the initial conditions. These bounds serve to control the
fluid entropy in the region around the star surface, by preventing numerical
errors from causing an entropy decrease and allowing heating only within a
reasonable range. To check that our results are not influenced by the choice of
these atmosphere fixing thresholds, we evolved a few comparison cases in which
we increased the densities $\rho_{\text{atmo}}^{\text{cut}}$ and
$\rho_{\text{atmo}}$ by a factor of 10. These comparison evolutions deviated
only slightly from the primary evolutions, confirming that our atmosphere
treatment does not strongly impact the neutron star simulations.

\subsection{Cowling neutron star in spherical symmetry}

We begin with 1D evolutions in spherical symmetry. For these simulations, we
rewrite the conservation law \eqref{eq:conslaw} and the relativistic Euler
equations in terms of spherical coordinates $\{r, \theta, \phi\}$. The DG
formulation takes a form similar to \eqref{eq:dgfinal} in one dimension, but
with a spherical divergence $\partial_r (r^2 u^r) / r^2$ replacing the
Cartesian divergence $\partial_x u^x$. The fluid equations pick up an
additional momentum source term:
$s(S_r) = s(S_x) + \alpha p (g^{rr} \partial_r g_{rr} + 2 / r)$. To avoid the
coordinate singularity at $r=0$, we set up a symmetric domain on the interval
$[-24,24]$ and use a staggered grid so that no nodes are located at the origin.

On this domain, we consider three grids with different resolutions. The first
two, which we name \dii{} and \diii{}, have comparable resolutions to the grids
of our 3D simulations. These two grids differ in the order of the DG elements
near the surface of the star; linear elements are used in \dii{} vs quadratic
elements in \diii{}. The third grid, \diir{},%
\footnote{The grid names are structured as follows: the first letter encodes
the domain's topology (``{\tt I}'': interval; ``{\tt B}'': ball), the integer
gives the (radial) order of approximation of the elements near the surface of
the star, and a final ``{\tt R}'' indicates a refined, higher-resolution grid.}
has a higher resolution and is more aggressively refined around the surface of
the star. In all three grids, we divide the domain into five regions: the
interior of the star, the surface on the left/right, and the exterior on the
left/right. We use larger, higher-order elements in the interior and exterior
regions and smaller, lower-order elements in the neighborhood of the star's
surface. The number and order of the elements within each region are listed in
Table \ref{table:params1d}. We evolve the system until $t=10^4 \simeq 50$~ms.
On the lower-resolution grids \dii{} and \diii{}, we use time steps
$\Delta t = 0.04$ corresponding to $\Delta t / \Delta x_{\text{min}} = 0.29$.
On the higher-resolution grid \diir{}, we use time steps $\Delta t = 0.025$
with $\Delta t / \Delta x_{\text{min}} = 0.57$.

\begin{table}[tb]
\centering
\caption[The structure of the spherically symmetric NS grids \dii{}, \diii{},
  and \diir{}.]{
  The structure of the spherically symmetric NS grids \dii{}, \diii{}, and
  \diir{}. For each grid, the parameters defining the elements in the interior,
  right-side surface, and right-side exterior regions are given; the elements
  in the left-side surface and left-side exterior regions are obtained by
  symmetry. The interior and exterior regions of \diii{} are identical to those
  of \dii{}.
}
\label{table:params1d}
\begin{ruledtabular}
\begin{tabular}{ll|l|l|l|}
                         &                       & Extents       & $K_{\text{region}}$ & $N_{\text{region}}$ \\
\hline
\multirow{3}{*}{\dii{}}  & Interior              & $[-7.5, 7.5]$ & 25                  & 3                   \\
                         & Surface (right side)  & $[7.5, 10]$   & 10                  & 1                   \\
                         & Exterior (right side) & $[10, 24]$    & 7                   & 3                   \\
\hline
\multirow{1}{*}{\diii{}} & Surface (right side)  & $[7.5, 10]$   & 5                   & 2                   \\
\hline
\multirow{3}{*}{\diir{}} & Interior              & $[-8, 8]$     & 101                 & 3                   \\
                         & Surface (right side)  & $[8, 9]$      & 20                  & 1                   \\
                         & Exterior (right side) & $[9, 24]$     & 30                  & 3                   \\
\end{tabular}
\end{ruledtabular}
\end{table}

\begin{figure}[tb]
\centering
\includegraphics[width=\columnwidth]{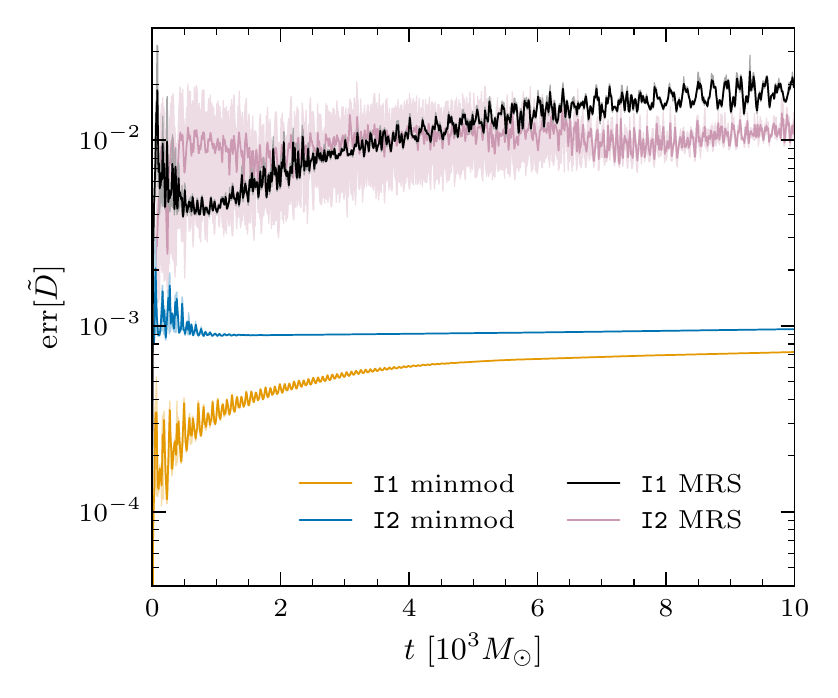}
  \caption[The density errors in the spherically symmetric Cowling NS
  evolution.]{
  The density errors in the spherically symmetric Cowling NS evolution. The
  four cases correspond to different choices of grid (\dii{} or \diii{}) and
  limiter (minmod or MRS) at the star surface. As some of the curves are highly
  oscillatory, we plot for each case the mean error in a solid line, and the
  envelope as a light-colored shaded region. The mean is computed by applying a
  Gaussian smoothing to the data, with half-width $\sigma \simeq 5$; the
  envelope minimum/maximum are computed in bins of width $\Delta \simeq 15$.
}
\label{fig:tovhy1dgridslims}
\end{figure}

We now compare evolutions of the spherically symmetric NS for different choices
of the grid and the limiter --- specifically the \dii{} or \diii{} grid, and
the minmod or MRS limiter. We plot in Fig.\ \ref{fig:tovhy1dgridslims} the
normalized density error $\text{err}[\tilde{D}]$ for each case over the
duration of the simulation. We first examine the two minmod cases. Here, the
data reveal two components in the dynamics: a short-period oscillatory behavior
and a gradual drift as the star settles to its numerical equilibrium
configuration on much longer timescales. The \diii{} case has a much higher
initial error and increased dissipation, as indicated by the more rapid decay
of the oscillatory component. The increased error and dissipation occur because
the minmod limiter linearizes the solution on the quadratic elements at the
surface, resulting in the loss of information. In the two MRS cases, we find
very different behavior: the density error is roughly an order of magnitude
larger (vs minmod) and grows over time, and the high-frequency oscillations are
not damped on the timescale of the simulation. Although the MRS evolutions are
stable (on this timescale), the star does not settle to an equilibrium.

\begin{figure}[tb]
\centering
\includegraphics[width=\columnwidth]{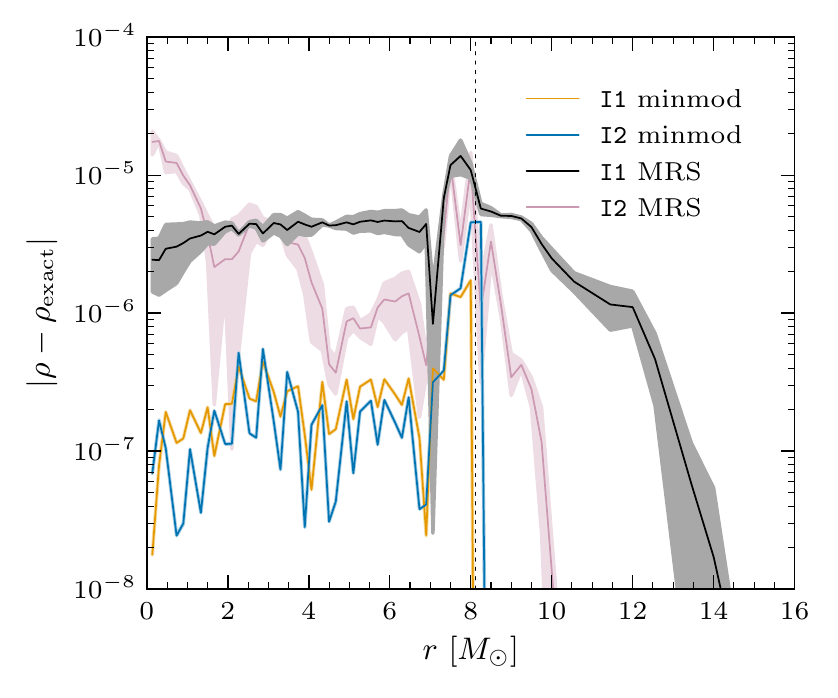}
  \caption[The rest-frame density error vs the radius, at time $t=2000$, in the
  spherically symmetric Cowling NS evolution.]{
  The rest-frame density error vs the radius, at time $t=2000$, in the
  spherically symmetric Cowling NS evolution. The shaded region shows the
  minimum/maximum errors, at each radius, from the left and right sides of the
  symmetric domain. We plot the mean error in the solid line. The vertical
  dotted line indicates the location of the TOV star surface at
  $r_{\text{NS}} \simeq 8.125$.
}
\label{fig:tovhy1dradial}
\end{figure}

To better understand this difference in behavior between the minmod and MRS
limiters, we now look at the distribution of the errors across the star. In
Fig.\ \ref{fig:tovhy1dradial}, we show the error in the fluid density vs the
stellar radius, at time $t=2000$. The solid lines in this plot show the
angle-averaged errors (i.e., the average of the left- and right-side data), and
the lighter filled region shows the spread in error values at fixed radius.
From this plot, we make two observations. First, while the minmod limiter
maintains excellent symmetry across the star, the MRS case shows a large spread
in the error values, indicating a loss of spherical (i.e., reflection)
symmetry. Second, while the density and velocity errors in the minmod case are
largest at the surface of the star, denoted by a vertical dotted line in the
figure, the fluid remains confined within the true surface of the star. When
using MRS, the star instead extends significantly beyond the true surface;
matter with non-negligible densities and large ($v > 0.01$) velocities is
present out to $r \simeq 15$. Our interpretation is that the MRS limiter
provides insufficient damping of small-scale fluctuations in the atmosphere
near the star.%
\footnote{Moe \emph{et al.}\ \cite{moe2015simple} discuss interpolating the
solution to a finer grid (to better sample its shape) before computing the
maxima and minima used by the limiter. We found no significant improvement in
behavior when using this interpolation.}
These slowly grow, leading to the expansion of the star beyond its true surface
and the contamination of the solution inside the star.

We conclude from these comparisons that the MRS limiter --- although effective
at handling shocks --- is poorly suited to the task of controlling a stellar
surface on the conforming grids that we are using. In the remainder of this
paper we therefore only show results obtained with the minmod limiter. We will
also restrict to results from grids with linear-order elements at the star
surface, as these are much less dissipative, given our use of this low-order
limiter.

In Sec.\ \ref{sec:tests}, we showed that our DG implementation has the expected
convergence properties for test problems with smooth solutions. For the NS
problem, the expected convergence behavior is less clear; the convergence rate
will be degraded by the discontinuity at the stellar surface, and, furthermore,
our use of a geometrically adapted grid with elements of different sizes and
different orders makes it more difficult to quantify the resolution. In spite
of this, we examine the convergence behavior with a series of short evolutions
in which we successively $h$ refine the \dii{} grid, and the time step, by
factors of 2 and 4 (in the interior region of the grid, we refine from 25 to
49, then 97, elements; this maintains a grid that straddles the origin). From
these evolutions, we measure the error in $\tilde{D}$ at time $t = 1000$; we
expect the error decrease to lie between a fourth-order convergence
(corresponding to the case where the error is dominated by the $N=3$ interior)
and first-order convergence (corresponding to the case where the error is
dominated by the discontinuity at the surface). We find (but do not show) that
as the grid is refined, the error decrease is consistent with a third-order
convergence.

\begin{figure}[tb]
\centering
\includegraphics[width=\columnwidth]{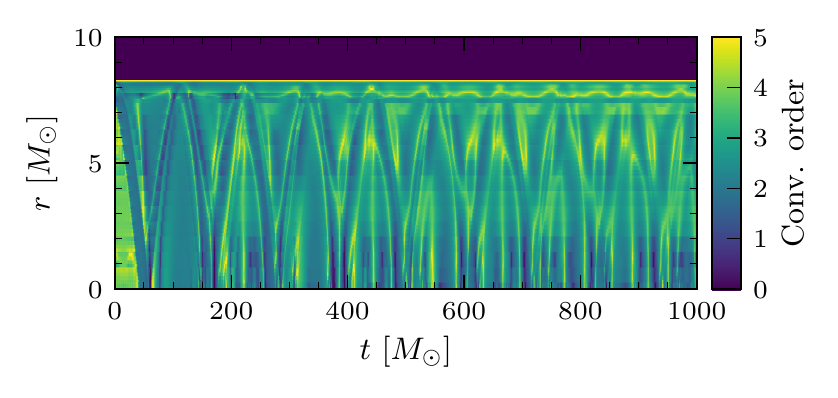}
  \caption[The local convergence order in the spherically symmetric Cowling
  NS evolution.]{
  The local convergence order in the spherically symmetric Cowling NS
  evolution. The convergence order is measured from the decrease in the
  element-average of err[$\tilde{D}$] between simulations on two computational
  grids: the \dii{} grid and a denser grid obtained by refining each \dii{}
  element into 4. Lighter colors correspond to a higher order of convergence,
  i.e., a more rapid decrease in error.
}
\label{fig:tovhy1dconvergence}
\end{figure}

To investigate the degradation in accuracy caused by the stellar surface, we
plot in Fig.\ \ref{fig:tovhy1dconvergence} the spatial and temporal variation
of a locally defined convergence order. At each point on the plot, the
convergence order is computed from Eq.\ \eqref{eq:convergencerelation} by
measuring the decrease in the element-averaged error in $\tilde{D}$ between
simulations on two computational grids: the \dii{} grid and the denser grid
obtained by a four times refinement of \dii{}. The interior of the star
initially shows the expected fourth-order convergence, but during the first 50
$M_{\odot}$ the order quickly decreases to roughly second order as
lower-accuracy data from the stellar surface propagate into the interior. We
find, in agreement with Refs.\ \cite{Bugner:2015gqa, Schoepe2017}, that as the
star settles to its numerical equilibrium, the order of convergence increases
again to roughly third-order convergence. This shows that the degradation in
convergence caused by the stellar surface is limited, and so the high-order
qualities of the DG method, to a large degree, continue to apply.

\begin{figure}[tb]
\centering
\includegraphics[width=\columnwidth]{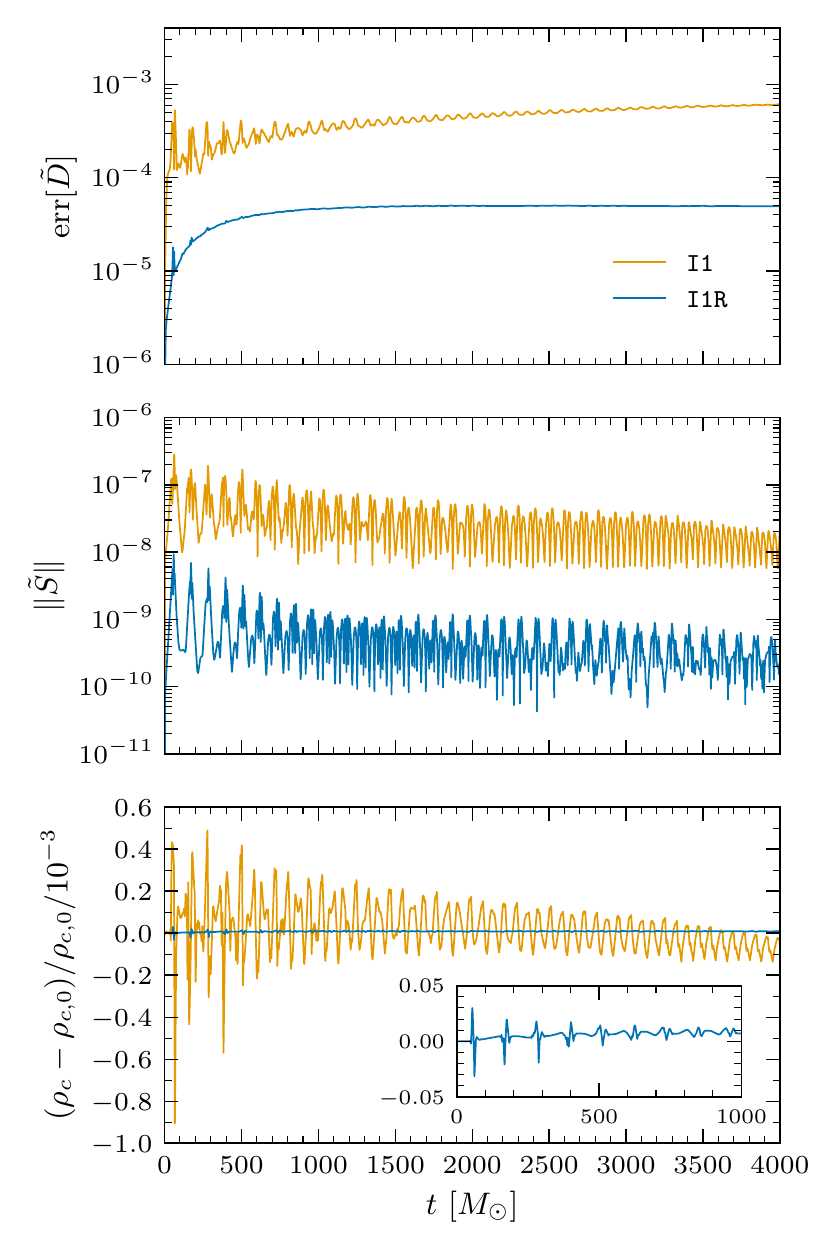}
\caption[The errors in the spherically symmetric Cowling NS evolution.]{
  The errors in the spherically symmetric Cowling NS evolution. The top
  (middle) panel shows the error in the conserved density $\tilde{D}$
  (conserved momentum $\tilde{S}_i$) for evolutions using the minmod limiter on
  the grids \dii{} and \diir{}. The bottom panel shows the evolution of the
  central density $\rho_c$ as a fractional error with respect to its initial
  value $\rho_{c,0}$. The inset in the bottom panel zooms in to better show the
  initial $\rho_c$ evolution in the \diir{} case; the \dii{} curve is omitted
  from the inset for visual clarity.
}
\label{fig:tovhy1dev}
\end{figure}

We now take a second, closer look at the spherically symmetric NS evolution. In
Fig.\ \ref{fig:tovhy1dev}, we compare evolutions of the spherically symmetric
NS on the \dii{} and \diir{} grids. We show, in the top two panels, the errors
in the conserved quantities $\tilde{D}$ and $\tilde{S}_r$ over the first
$4000 M_{\odot} \simeq 20$~ms of evolution time. The errors are lower by one or
two orders of magnitude in the \diir{} case. In the bottom panel, we plot the
time dependence of the central density $\rho_c$ as a fractional error with
respect to the initial central density $\rho_{c,0}$. We see in $\rho_c$ a
qualitative difference between the two evolutions, with the \diir{} case
showing a clearly periodic structure corresponding to the crossing time for
perturbations seeded at the surface of the star. In the full evolution to
$t = 10^4$, not shown in the figure, the high-resolution evolution maintains
its equilibrium, with the remaining oscillations in $\tilde{S}_r$ and $\rho_c$
slowly decaying. In the low-resolution evolution, the density error
$\text{err}[\tilde{D}]$ asymptotes to roughly $10^{-3}$, the oscillations in
$\norm*{\tilde{S}}$ slowly decay, and the central density continues to slowly
drop, reaching a $0.05\%$ deficit at $t = 10^4$.

While the DG method is fundamentally conservative, we have discussed (in Sec.\
\ref{sec:limiters}) how our use of limiters on deformed elements can violate
this property. The spherically symmetric simulations are also affected, even
though the 1D elements are themselves undeformed, because the limiter does not
account for the spherical Jacobian $4 \pi r^2$ that takes the 1D volume element
to the spherical volume element. The corrections to restore conservation
explored in Sec.\ \ref{sec:limiters} are no more effective in the 1D case;
conservation is restored at the expense of stability near the star surface. We
quantify the conservation error by tracking the NS's baryon mass
$M_b \equiv \int \tilde{D} 4 \pi r^2 dr$ during the evolution, as this should
be a conserved quantity. We find (but do not show) that $M_b$ slowly grows. On
the \dii{} grid, the relative error in $M_b$ (with respect to its initial
value) reaches roughly $10^{-4}$ at $t=10^4$, with over 90\% of this growth
occurring over the initial $4000 M_{\odot}$ as the star settles toward
equilibrium. On the \diir{} grid, the error grows to about $4\times10^{-7}$,
again mostly over the initial portion of the evolution.

We now reconsider Fig.\ \ref{fig:tovhy1dev}, and focus on the oscillatory
behavior seen in the different quantities. These oscillations are triggered by
errors from two sources: truncation errors from the evaluation of the exact TOV
solution on the finite-resolution numerical grid and the action of the limiter
which modifies the initial solution near the star's surface. These errors seed
perturbations of the various radial eigenmodes of the star, each of which
subsequently resonates with its corresponding eigenfrequency. A common test of
NS evolution codes is to compare the frequency spectrum of the simulated star
against the eigenfrequencies obtained from linear theory.

\begin{figure}[tb]
\centering
\includegraphics[width=\columnwidth]{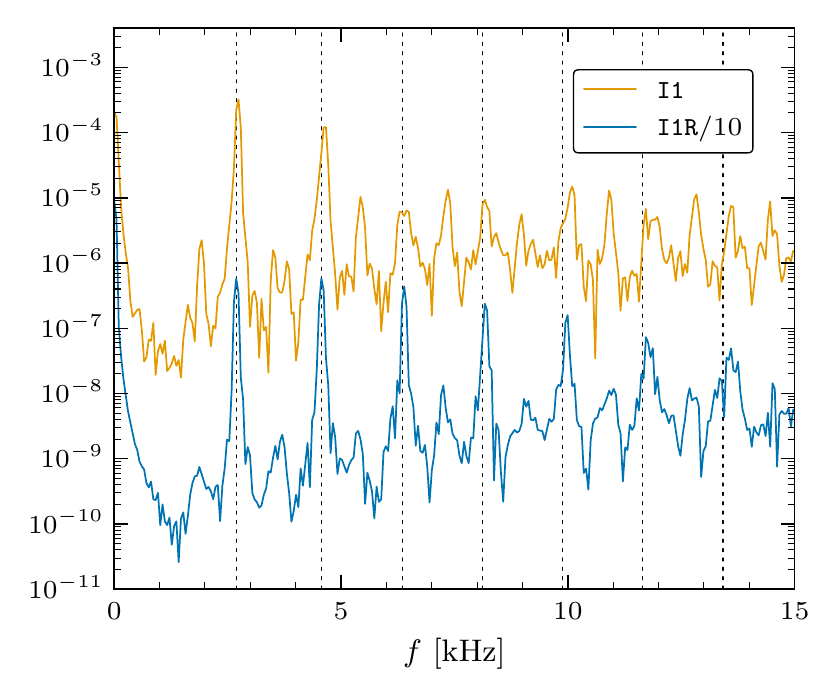}
\caption[The Fourier transform of the central rest-mass density $\rho_c$ from
  the spherically symmetric Cowling NS evolutions.]{
  The Fourier transform of the central rest-mass density $\rho_c$ from the
  spherically symmetric Cowling NS evolutions. The data from evolutions on the
  \dii{} and \diir{} grids are shown --- the \diir{} curve is shifted downward
  on the plot, by a factor of 10, for visual clarity. The vertical dotted lines
  indicate the frequencies of the fundamental normal mode and the first six
  harmonics. The units of the vertical axis are arbitrary.
}
\label{fig:tovhy1dfreq}
\end{figure}

To make this comparison, we compute the frequency spectrum from the central
rest-mass density during the first $4000 M_{\odot}$ of evolution time. After
subtracting the initial density offset $\rho_{c,0}$, we apply a Hanning window
to the time interval and compute the discrete Fourier transform. We plot in
Fig.\ \ref{fig:tovhy1dfreq} the absolute value of the Fourier coefficients
against frequency. The dotted vertical lines indicate the (Cowling) NS's radial
eigenmode frequencies, as listed in Table I of Ref.\
\cite{2002PhRvD..65h4024F}. The evolution on the \dii{} grid resolves few of
the star's eigenmodes; the spectrum has sharp peaks corresponding to the
fundamental mode and the first harmonic only. Modes with higher frequencies
(i.e., shorter wavelengths) are not spatially resolved by this computational
grid, and so the power they contain aliases into the lower-frequency modes. The
evolution on \diir{}, on the other hand, reproduces very clearly the
fundamental mode frequency and the first four harmonic frequencies. At higher
frequencies, the peaks are still identifiable, though they become broader and
less precisely centered. We note the presence of intermediate peaks in the
spectrum, shaped in a manner suggestive of sidebands; however, these features
are too noisy for unambiguous identification.

The evolution on \diir{}, with roughly 210 points across the NS's radius, has a
similar resolution to the 75-element case presented by Radice and Rezzolla
\cite{Radice:2011qr}. While these two simulations are not directly comparable
(the 1D star in the cited work self-consistently treats the gravity and uses a
uniform grid), we see a qualitative agreement in the number of resolved modes
and the precision with which they are resolved.

We conclude the discussion of 1D evolutions by noting that, while the \diir{}
grid has significantly reduced error, the lower-resolution \dii{} grid,
representative of the 3D resolution, is sufficient to resolve the important
features in the evolution. The lower-resolution case remains stable on long
timescales, and the oscillations as the star settles to its numerical
equilibrium correctly reflect the low-frequency eigenmodes from linearized
theory.

\subsection{Cowling neutron star in three dimensions}

\begin{figure}[tb]
\centering
\includegraphics[width=\columnwidth]{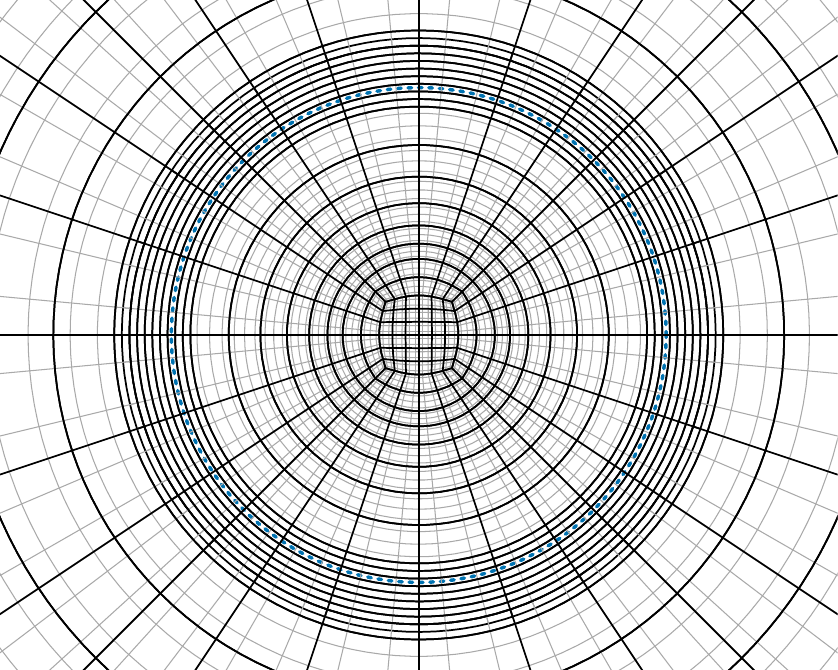}
\caption[The \dbi{} grid used in the 3D NS evolutions.]{
  The \dbi{} grid used in the 3D NS evolutions. The thick dotted circle
  indicates the location of the star's surface. The black lines show the
  element boundaries, and the light grey lines show the Gauss-Legendre-Lobatto
  grid within each element. The details of the grid mappings and structure are
  given in Appendixes \ref{app:cubedsphere} and \ref{app:cubedsphereNS},,
  respectively.
}
\label{fig:nsdomains}
\end{figure}

The simulation domain for the 3D star is a filled ball extending to
$r_{\text{max}} = 24$. We consider two different cubed-sphere grids on this
domain, constructed using the mappings detailed in Appendix
\ref{app:cubedsphere}. As in the spherically symmetric case, we adapt these
grids to the geometry by using larger, higher-order elements in the star's
interior as well as outside the star. In the region near the surface, the grids
are composed of thin cubed-sphere shells with a linear basis in the radial
direction. The first grid, \dbi{}, has a similar resolution across the radius
of the star to the \dii{} grid used in the 1D evolutions; an equatorial cut
through this grid is shown in Fig.\ \ref{fig:nsdomains}. The second grid,
\dbir{}, is obtained by adaptively refining \dbi{}: the large elements in the
interior and exterior regions are $p$-refined and the thin elements near the
surface of the star are $h$-refined, i.e., are radially split into thinner
shells. The complete description of these two grid structures is given in
Appendix \ref{app:cubedsphereNS}. In these 3D evolutions, we apply as before
the minmod limiter to the surface region of the grid only. We evolve the
hydrodynamics system until $t = 10^4 \simeq 50$ ms, with time steps
$\Delta t = 0.04$ on the \dbi{} grid
($\Delta t / \Delta x_{\text{min}} \simeq 0.61$) and $\Delta t = 0.025$ on the
\dbir{} grid ($\Delta t / \Delta x_{\text{min}} \simeq 0.59$).

When evolving the star on these cubed-sphere grids, we find a numerical
instability in the conserved momentum $\tilde{S}_i$ that leads to an
exponential growth of this quantity on $\mathcal{O}(100 M_{\odot})$ timescales.
This numerical instability is caused by the aliasing of the spectral modes as a
result of an insufficiently resolved quadrature rule in the DG method [see the
paragraph below \eqref{eq:dgint}]; we therefore filter the $\tilde{S}_i$
variable. In the central (as described in Appendix \ref{app:cubedsphereNS})
portion of the grid, the filter takes the form \eqref{eq:filter} with
$\alpha = 36$ and $s=6$.%
\footnote{With these values, the filter reduces power in approximately the
upper half of the modes. The power in the highest mode is reduced to round
off.}
In the cubed-sphere shells that make up the bulk of the interior, the
numerical instability is weaker and is controlled by a milder filter with
$\alpha = 36$ and $s=12$. With these filters, the NS evolutions remain stable
until at least $t = 10^4$. As the stars presented in this paper have a rest
state with no velocity, i.e., $\tilde{S}_i=0$, with dynamics that consist
primarily of short-timescale oscillations while the system settles to the rest
state, the filters have only a minor effect on the long-term evolution. For
stars undergoing rotation or pronounced dynamics, the filtering should not
qualitatively affect the results but would reduce the method's order of
convergence.

\begin{figure}[tb]
\centering
\includegraphics[width=\columnwidth]{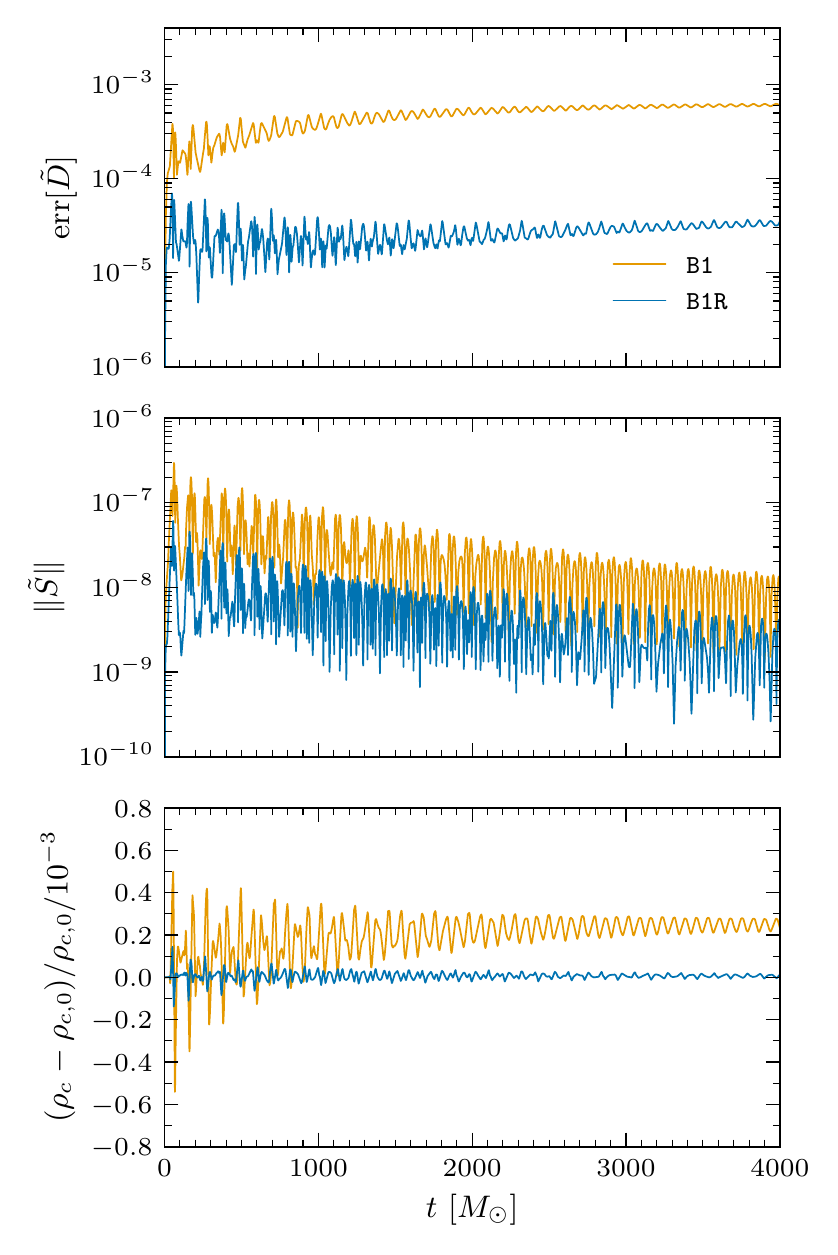}
\caption[The errors in the 3D Cowling NS evolution.]{
  The errors in the 3D Cowling NS evolution. The top (middle) panel shows the
  error in the conserved density $\tilde{D}$ (conserved momentum $\tilde{S}_i$)
  for evolutions using the minmod limiter on the grids \dbi{} and \dbir{}. The
  bottom panel shows the evolution of the central density $\rho_c$ as a
  fractional error with respect to its initial value $\rho_{c,0}$.
}
\label{fig:tovhyev}
\end{figure}

We plot, as before, the errors in $\tilde{D}$, $\tilde{S}_i$, and $\rho_c$
during the first $4000 M_{\odot}$ of evolution time in Fig.\ \ref{fig:tovhyev}.
The errors in the \dbi{} simulation closely match those seen in the spherically
symmetric case (cf.\ the \dii{} curves in Fig.\ \ref{fig:tovhy1dev}) --- this
is expected, given the comparable resolution and the spherically symmetric
nature of the problem. We note that the gradual decrease in central density
seen in the 1D simulation is not observed in three dimensions, and the central
density instead approaches a constant as the star settles. The errors in the
\dbir{} case are reduced by about an order of magnitude. Note that this case
cannot be directly compared to the high-resolution 1D case \diir{}, which has a
substantially higher resolution across the NS radius. In the full evolution to
$t = 10^4$, for both grids, the oscillations damp away, and the errors tend
toward a constant equilibrium value. We again check the conservation error by
tracking the baryon mass, which we find to slowly grow during the evolution.
The relative error in $M_b$ in the \dbi{} evolution is comparable to the \dii{}
case, reaching about $10^{-4}$ at $t = 10^4$. For the \dbir{} case the relative
error is about $6\times10^{-6}$. Again, most of these drifts are accumulated
during the initial settling of the star.

\begin{figure}[tb]
\centering
\includegraphics[width=\columnwidth]{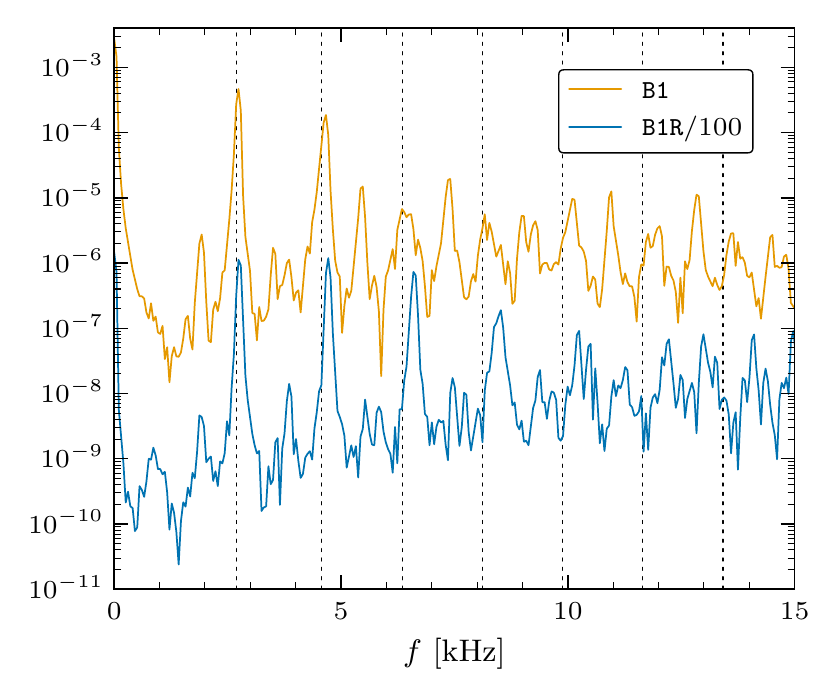}
\caption[The Fourier transform of the central rest-mass density $\rho_c$ from
  the 3D Cowling NS evolution.]{
  The Fourier transform of the central rest-mass density $\rho_c$ from the 3D
  Cowling NS evolution. The data from evolutions on the \dbi{} and \dbir{}
  grids are shown --- the \dbir{} curve is shifted downward on the plot, by a
  factor of 100, for visual clarity. The vertical dotted lines indicate the
  frequencies of the fundamental normal mode and the first six harmonics. The
  units of the vertical axis are arbitrary.
}
\label{fig:tovhyfreq}
\end{figure}

We compute the frequency spectrum of the stellar oscillations from $\rho_c$,
using the procedure described for the spherically symmetric case. The results
are shown in Fig.\ \ref{fig:tovhyfreq}. Comparing the \dbi{} spectrum to the
\dii{} spectrum from Fig.\ \ref{fig:tovhy1dfreq}, we see good agreement: the
first two resonant frequencies are clearly resolved, and additional peaks at
higher frequencies are suggestive but not conclusive. Going from \dbi{} to
\dbir{} we see improvement in the mode resolution, with a third and fourth peak
appearing in the frequency spectrum. These new peaks are increasingly shifted
toward higher frequencies, which indicates that the corresponding modes are
not yet fully resolved. The intermediate peaks seen in one dimension are still
visible in three dimensions but remain close to the noise level.

We also performed (but do not show) simulations of the 3D Cowling NS using the
same grid and limiter configurations that, in the 1D study, were found to be
problematic. These configurations use a grid where the surface elements have a
quadratic radial basis, and/or use the MRS limiter near the NS surface. For the
grid check, we employed a third cubed-sphere grid on the domain, \dbii{}, that
is similar to \dbi{} but uses thicker shells with a quadratic radial basis
(comparable to the \diii{} grid in one dimension; see Appendix
\ref{app:cubedsphereNS} for details). We found, as in the 1D case, that
evolutions on this grid using the low-order minmod limiter are stable over long
timescales, but with high dissipation and increased error. For the MRS limiter
check, we found, in contrast to the 1D case, that the 3D simulations are
unstable on $\mathcal{O}(1000 M_{\odot})$ timescales; the high-frequency
oscillations seen in the 1D case grow in 3D, presumably due to the limiter's
inability to control the additional tangential basis modes, and the star
rapidly becomes unstable.

\subsection{GR-hydro neutron star}

For the coupled GR-hydro evolutions, we again use the two grids \dbi{} and
\dbir{} from above. The hydrodynamics are treated as for the Cowling star, with
a minmod limiter at the star surface. We additionally evolve the spacetime
geometry, with the constraint damping parameters set to
\begin{align}
\gamma_0 &= 0.1 \exp[-(r/12)^2/2] + 0.01
\\
\gamma_1 &= -1
\\
\gamma_2 &= 3 \exp[-(r/12)^2/2] + 0.01.
\end{align}
The gauge function $H_{\sigma}$ is computed, as for the Kerr BH evolution, from
the contraction of the Christoffel symbols of the exact metric; it is constant
in time. We evolve the combined system until $t = 10^4 \simeq 50$ ms, with time
steps $\Delta t = 0.04$ on the \dbi{} grid
($\Delta t / \Delta x_{\text{min}} \simeq 0.61$) and $\Delta t = 0.025$ on the
\dbir{} grid ($\Delta t / \Delta x_{\text{min}} \simeq 0.59$).

\begin{figure}[tb]
\centering
\includegraphics[width=\columnwidth]{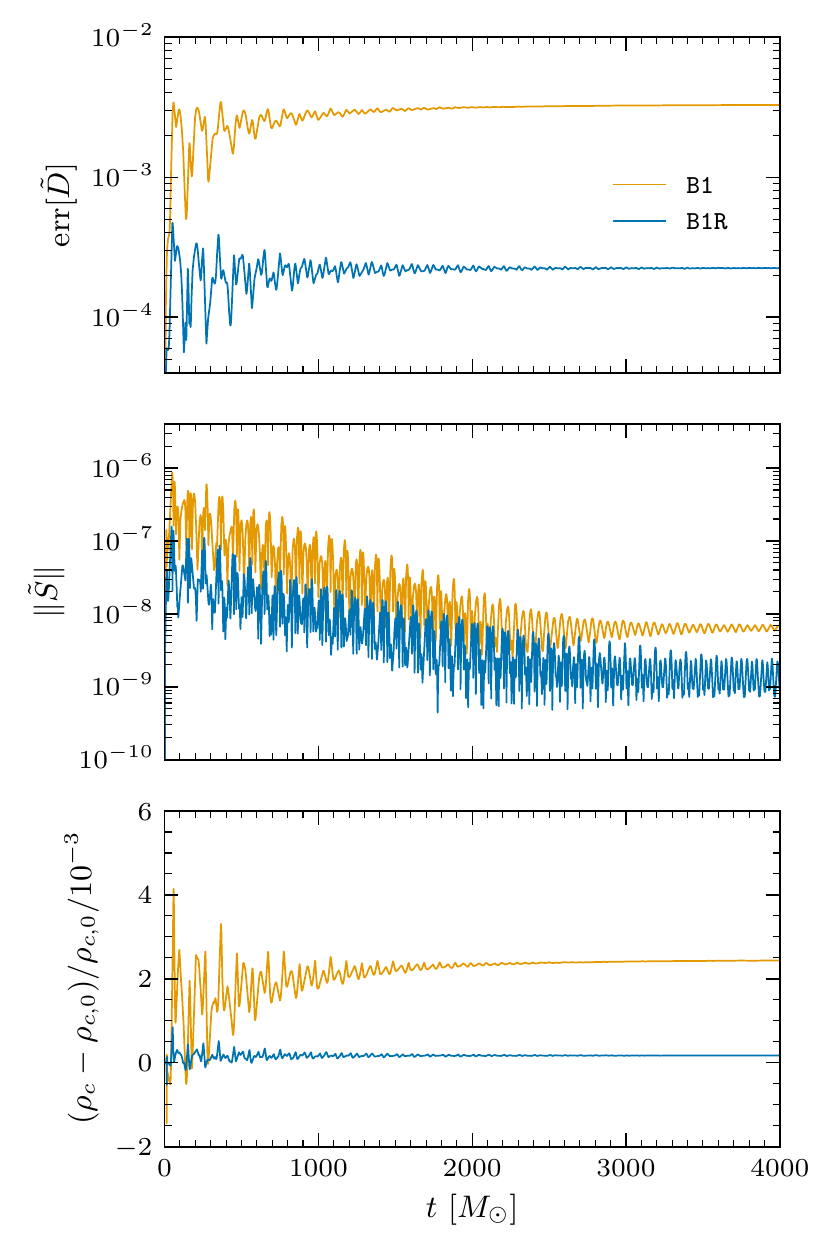}
\caption[The errors in the coupled GR-hydro NS evolution.]{
  The errors in the coupled GR-hydro NS evolution. The top (middle) panel shows
  the error in the conserved density $\tilde{D}$ (conserved momentum
  $\tilde{S}_i$) for evolutions using the minmod limiter on the grids \dbi{}
  and \dbir{}. The bottom panel shows the evolution of the central density
  $\rho_c$ as a fractional error with respect to its initial value
  $\rho_{c,0}$.
}
\label{fig:tovev}
\end{figure}

We show in Fig.\ \ref{fig:tovev} the errors in $\tilde{D}$, $\tilde{S}_i$, and
$\rho_c$ for the self-consistent NS evolution. Comparing the results from the
\dbi{} grid to the Cowling results of Fig.\ \ref{fig:tovhyev}, we see that the
self-consistent NS is more dissipative than the Cowling one --- the
oscillations decay quickly and become negligible by $t \sim 3000$.
Additionally, we see that the star settles to a different equilibrium, because
the gravity responds to the fluid rather than providing a fixed potential well.
The equilibrium central density is higher than the TOV value, indicating that
in its numerical equilibrium, the star has compressed slightly. The errors
using the higher-resolution grid \dbir{} are, as in the Cowling case,
significantly reduced as compared to the \dbi{} grid. In the full evolution to
$t = 10^4$, the \dbir{} case exhibits a slowly growing error component at late
times: from $t \simeq 7000$ onward, the errors increase by order 10\%. This
growing error is consistent with a weak numerical instability over
$\mathcal{O}(10^4 M_{\odot})$ timescales and could presumably be addressed by
improved filtering. We again check the conservation error by tracking the
baryon mass during the evolution. The relative error in $M_b$ for the \dbi{}
case is about $10^{-4}$ at $t = 10^4$, as in the Cowling case, and as before is
mostly accumulated during the initial settling of the star. In the \dbir{}
case, however, the relative error reaches about $2\times10^{-5}$, or three
times the value from the Cowling case. Here, the error is accumulated in two
phases: first during the initial settling of the star and again at the end of
the evolution when the slowly growing errors begin to affect the computation of
$M_b$.

\begin{figure}[tb]
\centering
\includegraphics[width=\columnwidth]{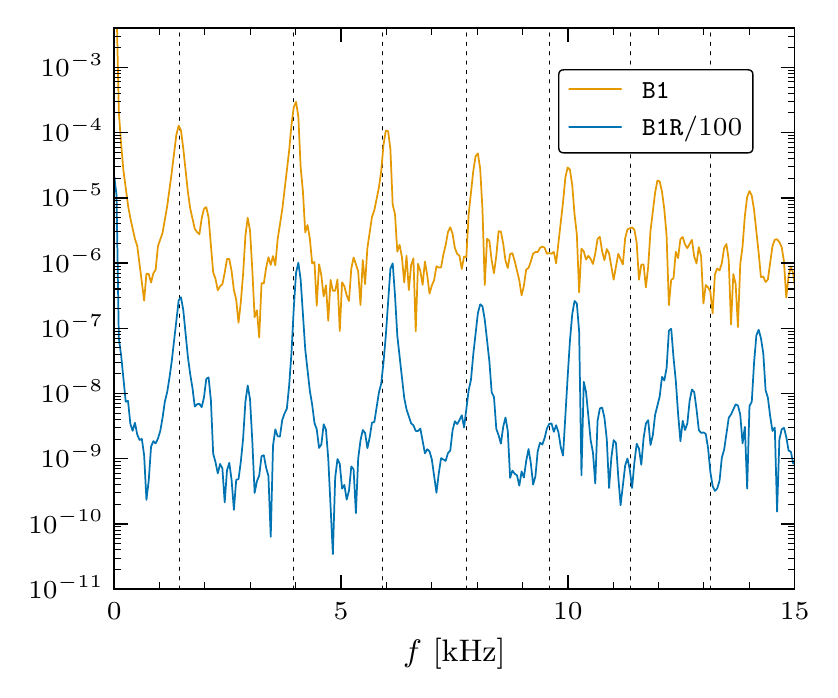}
\caption[The Fourier transform of the central rest-mass density $\rho_c$ from
  the coupled GR-hydro NS evolutions.]{
  The Fourier transform of the central rest-mass density $\rho_c$ from the
  coupled GR-hydro NS evolutions. Results are shown for evolutions on the
  \dbi{} and \dbir{} grids --- the \dbir{} curve is shifted downward on the
  plot, by a factor of 100, for visual clarity. The vertical dotted lines
  indicate the frequencies of the fundamental normal mode and the first six
  harmonics. The units of the vertical axis are arbitrary.
}
\label{fig:tovfreq}
\end{figure}

We compute once more the frequency spectrum of the stellar oscillations from
$\rho_c$, and we show in Fig.\ \ref{fig:tovfreq} the results from the
evolutions on both grids. We also indicate the first seven radial eigenmode
frequencies from linear theory%
\footnote{These eigenfrequencies were kindly provided to us by David Radice;
see the discussion of Fig.\ 11 of Ref.\ \cite{Radice:2011qr} for details.}
by the vertical dotted lines. In the lower-resolution \dbi{} case, we see clear
peaks in the spectrum from the fundamental mode up through the sixth harmonic.
The first three of these peaks are sharpest, indicating well-resolved modes;
the subsequent peaks become gradually less prominent and increasingly shifted
toward higher frequencies. The \dbir{} case is qualitatively similar --- we see
the same seven peaks in the spectrum, and although they are more prominent than
in the \dbi{} case because the noise floor is lower, the shift toward high
frequencies persists. Compared to the Cowling case, more modes are resolved. We
also note that the intermediate peaks seen in the Cowling case are no longer
prominent in the full GR-hydro results.

We conclude our analysis by comparing the accuracy of the DG and FV methods for
the NS problem. We use the \textsc{SpEC} GR-hydro code --- a FV code that takes
a dual-grid approach for coupled GR-hydro problems --- to perform additional
evolutions of the NS. The spacetime is evolved on a high-resolution grid of
nested spherical shells using a pseudospectral penalty method, closely related
to the DG method presented in this paper. The matter is evolved on a Cartesian
grid covering the interval $[0,12]$ in each direction (octant symmetry is
imposed), using a fourth-order finite-difference scheme with a WENO
reconstructor. We consider two resolutions for the FV grid. For the base
resolution, we require that the FV grid has the same number of grid points
within the volume of the star as the \dbi{} grid of the DG evolution. This
corresponds to a grid of $51^3$ points on $[0,12]^3$. The high-resolution grid
uses $101^3$ points --- far more than \dbir{}. These two cases are labeled
\dfv{} and \dfvr{} respectively.

\begin{figure}[tb]
\centering
\includegraphics[width=\columnwidth]{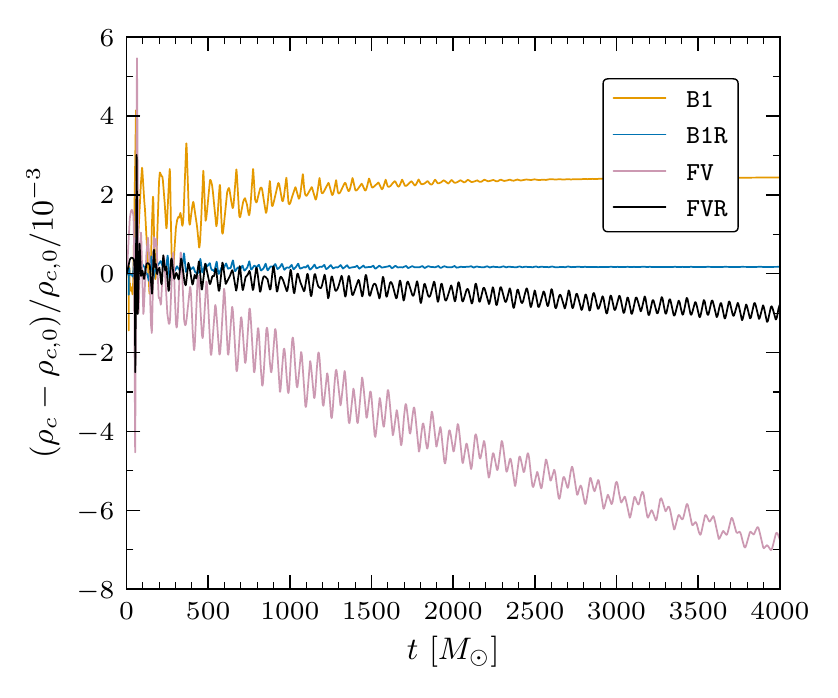}
\caption[The central-density error in the coupled GR-hydro NS, for
  evolutions using the DG and FV methods.]{
  The central-density error in the coupled GR-hydro NS, for evolutions using
  the DG and FV methods. For each method, two resolutions are shown: the DG
  method uses the grids \dbi{} (base) and \dbir{} (refined, with two times as
  many grid points), and the FV method uses the grids \dfv{} (base) and \dfvr{}
  (refined, with eight times as many grid points).
}
\label{fig:tovspechy}
\end{figure}

In Fig.\ \ref{fig:tovspechy}, we compare the central density errors in
evolutions with the DG and FV methods. The DG results make use of the grids
\dbi{} and \dbir{} (a two times increase in the number of grid points), and the
FV results make use of \dfv{} and \dfvr{} (an eight times increase) described
above. Comparing the results, we find a few differences between the DG and FV
evolutions. First, the DG method is more dissipative than the FV method used,
with the star's oscillations damping away by $t \sim 3000$. A contributing
factor to the increased dissipation is the use of a low-order shock-capturing
scheme in the surface regions for the DG evolutions vs the high-order
reconstruction scheme of the FV method. Second, the error in the central
density is greatly reduced in the DG evolution, primarily because of the
negligible drift rate after the star has settled to its numerical equilibrium.
Finally, in going to the refined \dbir{} and \dfvr{} grids, we find that the
error decreases more rapidly in the DG case, even though the resolution change
is smaller. This is because the DG method has higher order in the bulk of the
star's interior, so that $p$-refinement leads to rapid convergence. Precise
statements about the order of convergence for the DG results are difficult to
make, however, because we use geometrically adapted grids with elements of
different order.

\section{Conclusions}
\label{sec:conclusion}

In this paper, we have presented 3D evolutions using a DG method of (a) a Kerr
BH and (b) a general-relativistic NS treated self-consistently. We adopted the
DG formulation of Teukolsky \cite{teukolsky2015} to solve the generalized
harmonic formulation of Einstein's equations and the Val\`{e}ncia formulation
of general-relativistic hydrodynamics. We used conforming grids to take
advantage of the problem geometries, and we evolved the spacetime and matter
simultaneously on these grids. We implemented the DG method in the
\textsc{SpEC} framework and showed convergence and shock-capturing tests for
our code. We also evolved a NS under the Cowling approximation (fixed spacetime
metric) in spherical symmetry and in three dimensions.

With the 3D Kerr BH evolution, we showed that the DG method is accurate and
stable for long-timescale spacetime evolutions. By adapting the grid to the
(nearly) spherical geometry of the BH spacetime, we were able to excise the
singularity from the domain --- a promising result for the future use of the DG
method in compact-object binary simulations. The success of the DG method here
draws on previous successes of the (closely related) spectral penalty method
for the BH problem.

For the NS, we again showed long and stable evolutions, and we additionally
recovered the eigenfrequencies from linearized theory. By using domains
conforming to the star's spherical geometry and adapted to resolve the surface,
we were able to obtain good accuracy with comparatively few elements and a
low-order shock-capturing scheme. We compared the DG evolution to a FV
evolution and found significantly lower errors and an improved rate of
convergence from the DG case.

One of the advantages of the DG method over the FV method is that it is easier
to scale the algorithm on large machines. However, we were not able to show
scaling results from our implementation in \textsc{SpEC}. As discussed, the
\textsc{SpEC} framework scales poorly to large numbers of elements. For the NS
results shown, the domains are composed of over 5000 elements, enough for
\textsc{SpEC}'s scaling to break down and for timing measurements to lose their
significance. We do note that our DG method, which uses the same grid for the
spacetime geometry and the matter, solves the Einstein equations on a denser
grid of points than the dual-grid \textsc{SpEC} GR-hydro code. This adds a
significant computational cost for the runs presented in this paper. The added
cost would be reduced in the context of a science-producing simulation with a
spacetime grid extending to large radii, as the addition of some extra grid
points in the central portion of the domain would be less significant.

Improvements to our work will include the adoption of higher-order
shock-capturing schemes (e.g., WENO) to lower the errors in the treatment of
the star surface. The development of an adaptive mesh-refinement scheme will
allow geometrically adapted grids to be used in systems with reduced symmetry
and/or dynamics. These improvements are planned for implementation in the
\textsc{SpECTRE} code, where they will enable evolutions with the DG method of
dynamical systems such as rotating or dynamically unstable stars.

\begin{acknowledgments}
We thank Andy Bohn, Mike Boyle, Nils Deppe, Matt Duez, Francois Foucart, Jan
Hesthaven, Curran Muhlberger, and Will Throwe for many helpful conversations
through the course of this work. We gratefully acknowledge support for this
research from the Sherman Fairchild Foundation; from NSF Grants
No.\ PHY-1606654 and No.\ AST-1333129 at Cornell; and from NSF Grants
No.\ PHY-1404569, No.\ PHY-1708212, and No.\ PHY-1708213 at Caltech. F.H.\
acknowledges support by the NSF Graduate Research Fellowship under Grant
No.\ DGE-1144153. Computations were performed at Caltech on the Zwicky cluster,
which is supported by the Sherman Fairchild Foundation and by NSF Grant
No.\ PHY-0960291, and on the Wheeler cluster, which is supported by the Sherman
Fairchild Foundation.
\end{acknowledgments}

\appendix

\section{Cubed-sphere mappings}
\label{app:cubedsphere}

In simulations of systems with spherical geometries, we use grids based on the
cubed sphere \cite{Ronchi1996}. The cubed sphere is obtained by projecting the
faces of a cube onto its circumscribed sphere, thereby defining a grid on the
sphere composed of six deformed Cartesian grid patches. The radial direction is
introduced by a tensor product, giving a grid on a hollow spherical shell
composed of six mapped cubes; we call each of these mapped cubes a wedge of the
spherical shell. For our NS simulations, however, we require a filled ball
topology, rather than a hollow spherical shell.

\begin{figure}[tb]
\hspace*{\fill}
\subfloat[Undeformed cube]{
\includegraphics[width=0.36\columnwidth]{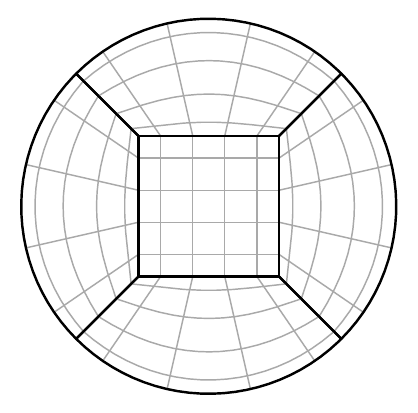}
}
\hfill
\subfloat[Rounded cube]{
\includegraphics[width=0.36\columnwidth]{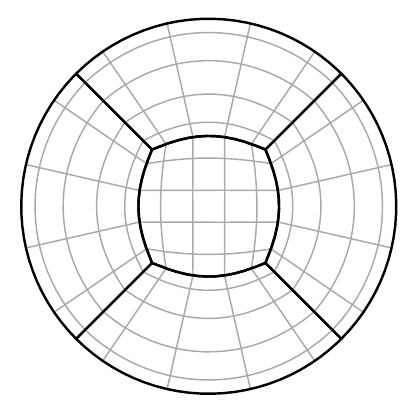}
}
\hspace*{\fill}
\caption[Two grids on the filled ball, constructed from a cubed sphere with
  (a) an undeformed cube as central element and (b) a rounded cube as central
  element.]{
  Two grids on the filled ball, constructed from a cubed sphere with (a) an
  undeformed cube as central element and (b) a rounded cube as central
  element. Both panels show an equatorial cut through the grid. The grids are
  obtained from the mappings described in Appendix \ref{app:cubedsphere}, with
  parameters $c_{\text{min}} = 0$ in panel (a) and $c_{\text{min}} = 0.66$
  in panel (b); in both panels, $c_{\text{max}} = 1$, $x_{\text{min}} = 0.75$,
  and $x_{\text{max}} = 2$. The black lines show the element boundaries, and
  the light grey lines show the Gauss-Legendre-Lobatto grid within each element
  for order $N=5$.
}
\label{fig:fillcubedsphere}
\end{figure}

To obtain a grid on the filled ball, one possibility is to insert a cube-shaped
element at the center of the grid and deform the inner surface of the spherical
shell so that it conforms to this cube. This is shown in panel (a) of Fig.\
\ref{fig:fillcubedsphere}. In numerical experiments, we find that this grid
configuration often suffers from large errors along the diagonal edges where
three of the wedges meet (e.g., the line $x=y=z$) because of the large grid
distortions at these locations. This source of error can be reduced by
inserting a ``rounded'' cube, which reduces the grid distortion in the wedges,
as shown in panel (b) of Fig.\ \ref{fig:fillcubedsphere}. As we are not aware
of previous uses of such a grid configuration, we show here the mappings used.

\subsection{Wedges}

The geometry of a cubed-sphere wedge is specified by its inner and outer
surfaces. Each of these surfaces is described by two parameters --- its
curvature $c$ and its position $x$. The surface's curvature $c \in [0,1]$
controls the shape: when $c=0$, the surface is flat (i.e., the six wedges
together form a cube), and when $c=1$, the surface is spherical. The surface's
position is the ``radius'' from the origin to the center of the surface, i.e.,
the point where the surface intersects the $x/y/z$ axis. The positions
$x_{\text{min}}$ and $x_{\text{max}}$ of the inner and outer surfaces satisfy
$0 < x_{\text{min}} < x_{\text{max}}$.

The mapping from the reference element to each cubed-sphere wedge is a radial
interpolation between the wedge's inner and outer surfaces, and is computed by
composing four transformations,
\begin{equation}
\label{eq:cubedspherewedgemap}
\textbf{x} (\bar{\textbf{x}}) = (
\textbf{x}_{\text{rot}}
\circ \textbf{x}_{\text{cs}}
\circ \textbf{x}_{\text{tan}}
\circ \textbf{x}_{\text{affine}}) (\bar{\textbf{x}}).
\end{equation}
The actions of these transformations are:
\begin{enumerate}
\item $\textbf{x}_{\text{affine}}$ shifts and scales the reference cube along
  the $+x$ axis to obtain a parallelepiped spanning
  $0 < x_{\text{min}} \le x \le x_{\text{max}}$. The $y$ and $z$ coordinates
  are unaffected.
\item $\textbf{x}_{\text{tan}}$ maps the tangential coordinates $y$ and $z$
  according to $y \to \tan(\pi y /4)$, and likewise for $z$. This shifts the
  grid point distribution tangentially inward to produce a more uniform,
  equiangular grid when the destination surface is spherical. This
  transformation is optional; we use it for the spherical-shell grids of the
  Kerr BH and spherical accretion tests, but elsewhere, we omit it.
\item $\textbf{x}_{\text{cs}}$ deforms the parallelepiped into one wedge of
  the cubed sphere, intersecting the $+x$ axis at $x_{\text{min}}$ and
  $x_{\text{max}}$. It is computed with the intermediate steps
\begin{align}
a &= 1/\sqrt{1 + \bar{y}^2 + \bar{z}^2}
\\
b_{\text{min}} &= x_{\text{min}} \left[1 + c_{\text{min}} ( a - 1 ) \right]
\\
b_{\text{max}} &= x_{\text{max}} \left[1 + c_{\text{max}} ( a - 1 ) \right]
\\
\xi &= b_{\text{min}} + (b_{\text{max}} - b_{\text{min}})\frac{\bar{x}
  - x_{\text{min}}}{x_{\text{max}} - x_{\text{min}}}
\\
\textbf{x}_{\text{cs}}(\bar{\textbf{x}}) &= (\xi, \xi \bar{y}, \xi \bar{z}).
\end{align}
\item $\textbf{x}_{\text{rot}}$ rotates the wedge to its location on the
  sphere, corresponding to one of the axes $+x$, $-x$, $+y$, $-y$, $+z$, or
  $-z$.
\end{enumerate}

Figure \ref{fig:fillcubedsphere} shows two (filled) cubed-sphere grids where
the outer surfaces are spherical and the inner surfaces have $c = 0$ and 0.66.
Figure \ref{fig:bhdomains} shows a cubed-sphere grid where both surfaces are
spherical, and each wedge is divided radially and tangentially into several
elements. This is achieved by dividing the unit cube into the corresponding
elements before applying the chain of maps in \eqref{eq:cubedspherewedgemap}.

\subsection{Rounded central cube}

The mapping from the reference element to the rounded central cube is chosen to
conform to the inner boundary of the cubed-sphere wedges. The cube is
therefore parametrized by the $x_{\text{min}}$ and $c_{\text{min}}$ that give
the inner boundary of the wedges and by whether the equiangular transformation
is applied. The mapping is again obtained by composition,
\begin{equation}
\label{eq:cubedspherecubemap}
\textbf{x} (\bar{\textbf{x}}) = (
\textbf{x}_{\text{rc}}
\circ \textbf{x}_{\text{tan}}) (\bar{\textbf{x}}),
\end{equation}
with $\textbf{x}_{\text{rc}}$, the transformation that deforms the cube,
given by
\begin{align}
a &= 1/\sqrt{1 + \bar{x}^2\bar{y}^2 + \bar{x}^2\bar{z}^2 + \bar{y}^2\bar{z}^2
  - \bar{x}^2\bar{y}^2\bar{z}^2}
\\
b_{\text{min}} &= x_{\text{min}} \left[1 + c_{\text{min}} ( a - 1 ) \right]
\\
\textbf{x}_{\text{rc}}(\bar{\textbf{x}})
  &= (b_{\text{min}} \bar{x}, b_{\text{min}} \bar{y}, b_{\text{min}} \bar{z}).
\end{align}
Inverting this mapping for
$\bar{\textbf{x}} = \textbf{x}_{\text{rc}}^{-1}(\textbf{x})$
requires root finding and is done numerically.

The right panel of Fig.\ \ref{fig:fillcubedsphere} shows a cubed-sphere grid
with a rounded central cube. Figure \ref{fig:nsdomains} shows a rounded cube as
used in the NS simulation grids; just as for the wedges, the division of the
central cube into several elements is achieved by dividing the unit cube prior
to applying the chain of maps in \eqref{eq:cubedspherecubemap}.

\section{Neutron star simulation grids}
\label{app:cubedsphereNS}

The simulation domain for the 3D NS evolutions is a filled ball extending to
$r_{\text{max}} = 24$. We use three different cubed-sphere grids on this
domain: \dbi{}, \dbii{}, and \dbir{}. Here, we define each of these grids.

The \dbi{} grid is shown in Fig.\ \ref{fig:nsdomains}. For the bulk of the
stellar interior, the grid is composed of nested, spherical cubed-sphere shells
containing higher-order elements. In the center of the domain, the grid
transitions to a rounded cube using the mappings described in Appendix
\ref{app:cubedsphere}. In the region near the surface, we use thinner elements
with fewer points; for \dbi{} there are ten shells of thickness
$\Delta r = 0.25$, each of which contains elements with a linear basis in the
radial direction (we denote this as $N_r=1$). Outside the star, the grid is
again made up of larger, higher-order elements. The details of this radial
structure are given in Table \ref{table:params3d}, which lists the parameters
of the cubed-sphere shells that make up the grid. The angular structure of
\dbi{} is obtained by tangentially splitting each wedge into $6\times6$
elements, each of which has a basis of order $N_{\text{tan}}=3$ in the two
tangential directions. The resolution of the central rounded cube is set by
conforming to the angular grid of the shell. The equiangular tangent mapping is
\emph{not} applied --- omitting this mapping gives a more optimal resolution of
the cube in the center of the star. The \dbi{} grid has a total of 5184
elements, with $\Delta x_{\text{min}} \simeq 0.0657$.

\begin{table}[tb]
\centering
\caption[The radial structure of the 3D NS grids, \dbi{}, \dbii{}, and
  \dbir{}.]{
  The radial structure of the 3D NS grids, \dbi{}, \dbii{}, and \dbir{}.
  For each region of each grid, the location and curvature of the surfaces that
  bound the cubed-sphere elements are given. Duplicated information is omitted:
  the unspecified regions of \dbii{} are identical to those of \dbi{}.}
\label{table:params3d}
\begin{ruledtabular}
\begin{tabular}{ll|l|l|l|}
                        &        & $x_i$                               & $c_i$         & $N_r$ \\
\hline
\multirow{4}{*}{\dbi{}} & Center & 1.3, 1.9, 2.5                       & 0.55, 0.85, 1 & 4     \\
                        & Int.\  & 2.5, 3.0, 3.6, 4.33, 5.2, 6.24, 7.5 & 1             & 4     \\
                        & Surf.\ & 7.5, 7.75, 8, ..., 9.5, 9.75, 10    & 1             & 1     \\
                        & Ext.\  & 10, 12, 15, 18, 21, 24              & 1             & 3     \\
\hline
\multirow{1}{*}{\dbii{}}& Surf.\ & 7.5, 8, 8.5, 9, 9.5, 10             & 1             & 2     \\
\hline
\multirow{4}{*}{\dbir{}}& Center & (see \dbi{})                        & (see \dbi{})  & 5     \\
                        & Int.\  & (see \dbi{})                        & (see \dbi{})  & 5     \\
                        & Surf.\ & 7.5, 7.625, 7.75, ..., 9.875, 10    & 1             & 1     \\
                        & Ext.\  & (see \dbi{})                        & (see \dbi{})  & 4     \\
\end{tabular}
\end{ruledtabular}
\end{table}

The \dbii{} grid differs from \dbi{} in the radial resolution of the surface
region. Where \dbi{} uses ten shells of linear order, \dbii{} instead uses five
spherical shells, of thickness $\Delta r = 0.5$, each of which contains
elements with a quadratic basis in the radial direction (i.e., $N_r=2$).
The \dbii{} grid has a total of 4104 elements, and the same
$\Delta x_{\text{min}}$ as \dbi{}.

The \dbir{} grid is obtained from \dbi{} by selectively refining to further
take advantage of the $hp$-adaptivity of the DG method: $h$-refinement is used
in the neighborhood of the surface where the solution is not smooth, and
$p$-refinement is used in the smooth interior and exterior regions. The radial
parameters are again given in Table \ref{table:params3d}; the angular
parameters are as for \dbi{} but with $N_{\text{tan}}=4$. This grid has 7344
elements, with $\Delta x_{\text{min}} \simeq 0.0447$, and has roughly twice as
many grid points inside the NS ($r \lesssim 8.125$) as the \dbi{} grid.

\section*{References}

%

\end{document}